\documentclass[10pt]{article}
\usepackage{amssymb,amsmath}
\usepackage{graphicx,xcolor}
\usepackage{mathrsfs}
\usepackage{hyperref}
\hypersetup{colorlinks=true,allcolors=[rgb]{0,0,0.6}}
\newtheorem{thm}{Theorem}[section]
\newtheorem{cor}[thm]{Corollary}
\newtheorem{lem}[thm]{Lemma}
\newtheorem{defn}[thm]{Definition}
\newtheorem{prop}[thm]{Proposition}
\newtheorem{remarks}[thm]{Remarks}
\def\proof{{\bf Proof. }}
\def\be{\begin{equation}}
\def\ee{\end{equation}}
\def\bea{\begin{eqnarray}}
\def\eea{\end{eqnarray}}
\def\bean{\begin{eqnarray*}}
\def\eean{\end{eqnarray*}}
\def\ea{\end{array}}
\def\ds{\displaystyle}

\def\Z{{\mathscr{Z}}}
\def\P{{\mathscr{P}}}
\def\L{{\mathscr{L}}}
\def\H{{\mathscr{H}}}
\def\M{{\mathscr{M}}}
\def\S{{\mathcal{S}}}
\def\D{{\mathcal{D}}}
\newcommand{\field}[1]{\mathbb{#1}}
\newcommand{\rz}{\field{R}}

\newcommand{\nz}{\field{N}}
\def\d{{\rm{d}}}
\def\ra{{\rangle}}
\def\la{{\langle}}

\def\fin{{$\hfill\square$}}
\def\hbarr{{\varepsilon}}

\def\Tr{{\rm{Tr}}}

\begin{document}
\title{On the rate of convergence for the mean field approximation of many-body quantum dynamics}
\author{Z.~Ammari$^*$, M.~Falconi\thanks{IRMAR, Universit{\'e} de Rennes I, UMR-CNRS 6625, Campus de Beaulieu, 35042 Rennes Cedex, France} , B.~Pawilowski\thanks{Fak. Mathematik, Univ. Wien, Oskar-Morgenstern-Platz 1, 1090 Wien, Austria
}}
\date{\today}

\maketitle
\begin{abstract}
We consider the time evolution of quantum states by  many-body Schr\"odinger dynamics and study the rate of convergence of their reduced density matrices in the mean field limit.  If the prepared state at initial time is of coherent or factorized  type and the number of particles $n$ is large enough then it is known that $1/n$ is the correct rate of convergence at any time. We show  in the simple case of bounded pair potentials that the previous rate of convergence holds in more general situations with possibly correlated prepared states. In particular, it turns out that the coherent structure at initial time is unessential and the important fact is
 rather the speed of convergence of all reduced density matrices of the prepared states. We illustrate our result with several numerical simulations and examples of multi-partite entangled quantum states borrowed from quantum information.
\end{abstract}

\tableofcontents
\section{Introduction}

The mean field theory provides in principle a fair  approximation of time evolved quantum  states by many-body  Schr\"odinger dynamics in the mean field scaling; namely when the number of particles is large and the pair interaction potential is proportionally weak. During the last decade, a strong activity around the mean-field problem has occurred within the community of mathematical physics. This in particular have led to a rigorous justification of the mean field approximation for singular potentials including Coulomb interaction as well as the derivation of the Gross-Pitaevskii  equation from many-body quantum dynamics  (see for instance \cite{AmBr,AmNi4,BGM,BEGMY,ChPa,ElSc,ErYa,ESY1,ESY2,FGS,FKP,FKS,GMP,KlMa,LSSY,Pi} and also \cite{GiVe1,GiVe2,Hep,Spo} for older results). More recently, emphasis has been placed on the speed of convergence of the mean-field approximation. This seems to be motivated by providing useful quantitative bounds and understanding higher order corrections (see \cite{Ana,C,CLS,F,GMM,KP,Pi,RoSc}).

The aim of our article, is to give at the level of a simple model more insight on the aforementioned problem. Actually, the rate of convergence is essentially  understood in the case of coherent or factorized type states with a particular structure. So, we can ask the following natural questions:
 \begin{itemize}
 \item  What should we expect if we start from another prepared state which is more correlated?
 \item Is the specific coherent structure of the known examples important?
 \item Can we determine the optimal rate of convergence in some examples?
 \item Does the rate of convergence improves under the effect of the quantum dynamics?
\end{itemize}
We will show that the rate of convergence at a given time depends essentially on the rate of convergence of all reduced density matrices of the prepared state at time $t=0$.
In fact, we are able to give a general condition on the prepared state that guaranties
a given speed of convergence at any time.  The  assumption we require at time zero, which is rather easy to check in initial  states,  is true at any time if it holds at $t=0$. This   allows  in particular to consider the question of improvement of the convergence over time while the question of optimality will be addressed through numerical analysis.

\bigskip
\noindent
 Consider for instance the many-body  Schr\"odinger Hamiltonian of an $n$-boson system
\bea
\label{schrod}
\mathbf{H}_n=\sum_{i=1}^n-\Delta_{x_i}+\frac{1}{n} \;\sum_{1\leq i<j\leq n} \, V(x_i-x_j)\;,
\eea
where $(x_1,\cdots,x_n)\in\mathbb{R}^{dn}$ and $V$ is  a real bounded potential satisfying $V(x)=V(-x)$. The self-adjoint operator $\mathbf{H}_n$ acts on the  space $L_s^{2}(\mathbb{R}^{dn})$ of symmetric square integrable functions. A function
$\Psi_n\in L^{2}(\mathbb{R}^{dn})$ is symmetric if $\Psi_n(x_1,\cdots,x_n)=\Psi_n(x_{\sigma_1},\dots,x_{\sigma_n})$
for any permutation $\sigma$ on the symmetric group $\mathfrak{S}(n)$. Suppose that
the system is in a prepared quantum state $\varrho_n$ at initial time $t=0$ (that is $\varrho_n$ is a non-negative trace class operator with $\Tr[\varrho_n]=1$). So, under the action of the  Schr\"odinger dynamics the system at time $t$ evolves into the state
$$
\varrho_n(t)=e^{i t \mathbf{H}_n}
\varrho_n e^{-i t \mathbf{H}_n}\,.
$$
The mean field approximation at the dynamical level is usually understood as the following picture: if the system is in an uncorrelated  state $\varrho_n=|\varphi^{\otimes n}\rangle\langle \varphi^{\otimes n}|$, with $||\varphi||_{L^2(\rz^d)}=1$, at initial time $t=0$ then it will evolve into a state close in some sense to an uncorrelated one $\varrho_n(t)\simeq|\varphi_t^{\otimes n}\rangle\langle \varphi_t^{\otimes n}|$ when $n$ is large and $\varphi_t$ is the solution of the nonlinear Hartree equation
\begin{eqnarray}
\label{hartree}
\left\{
  \begin{array}[c]{l}
    i\partial_t \varphi_t=-\Delta \varphi_t+(V*\left|\varphi_t\right|^{2})\varphi_t\,,\\
    \varphi_{t=0}=\varphi\,.
  \end{array}
\right.
\end{eqnarray}
 The above convergence is neither a strong nor a weak one but rather in the sense of reduced density matrices. More precisely, the convergence is understood as
\bean
\lim_{n\to\infty}\Tr[\varrho_n(t) A\otimes 1^{\otimes (n-p)} ]=
\langle\varphi_t^{\otimes p}, A \varphi_t^{\otimes p}\ra_{L^2(\mathbb{R}^{dp})}, \;\;
\eean
for any bounded (or compact) operator $A$ on $L^{2}(\mathbb{R}^{dp})$ and any $p\in\nz^*$ ($p$ is kept fixed while $n\to \infty$).\\
In some sense, the mean field approximation says essentially that the measurements
\bea
\label{correl}
\Tr[\varrho_n(t) \, A\otimes 1^{\otimes (n-p)}], \quad n\geq p\,,
\eea
for any observable $A$ on $L^2(\rz^{dp})$ converge, when $n$ goes to infinity while $p$ is kept fixed, to some classical or one particle quantities to be determined. Hence, the main quantities to be analyzed are the reduced density matrices of the time evolved states $\varrho_n(t)$. Recall that
for each $p\in\mathbb{N}^*$, the $p$-reduced density matrix of $\varrho_n(t)$
is the unique non-negative trace class operator $\varrho_n^{(p)}(t)$ on $L^2_s(\rz^{dp})$ satisfying
\bea
\label{eq.1}
\Tr[\varrho_n(t) \, A\otimes 1^{\otimes (n-p)}]=\Tr[ \varrho_n^{(p)}(t) \,A]\,,
\eea
for any bounded operator $A$ on $L^2(\rz^{dp})$. Therefore, the point is to determine
for each $p\in\mathbb{N}^*$ the limit and the rate of convergence of these quantities \eqref{eq.1} when the number of particles $n$ goes to infinity. It turns out that the limit  at $t=0$ may not exist and actually there is a difference between requiring convergence in \eqref{eq.1} for all bounded operators $A$ on $L^2(\rz^{dp})$, or convergence for compact operators only, since the weak and weak-$*$ topologies differ on the space of trace-class operators.
However, one can characterize all the limit points
 of $(\varrho_n^{(p)})_{n\geq p}$ with respect to the weak-$*$ topology in the space of trace-class operators (which is the dual space   of compact operators) and also
 describe their structure. Indeed,  at time $t=0$, we can show that there exists always a subsequence $(\varrho_{n_k})_{k\in\mathbb{N}}$ such that for each $p\in\mathbb{N}$, $1\leq p\leq n_k$, the reduced density matrices $(\varrho_{n_k}^{(p)})_{k\in\nz^*}$ converge  to   non-negative trace-class operators $\varrho_\infty^{(p)}$ in the weak-$*$ topology. Moreover, there exists a Borel probability measure $\mu$ on $L^2(\rz^d)$ such that
$$
\varrho_\infty^{(p)}=\int_{L^{2}(\rz^d)} |z^{\otimes p}\rangle\langle z^{\otimes p}| \, d\mu(z)\,.
$$
In this way we have characterized  all the possible limit points via subsequences of the reduced density matrices $(\varrho_n^{(p)})_{n\geq p}$ and identified  their structure. More details are given in Subsection \ref{sub.wigner} while here we summarize the main result in the proposition below.
We will use often the notation $\L^k(\mathfrak{h})$, $1\leq k\leq \infty$, to refer to the Schatten classes with $||\cdot ||_k$ denoting their norms.
\begin{prop}
\label{definetti}
Let  $(\varrho_{n})_{n\in\nz^*}$  be a sequence of density matrices with $\varrho_n\in\L^1(L^2_s(\rz^{dn}))$ for each $n\in\nz^*$. Suppose that for any $p\in\nz^*$ and  each compact operator $A\in\L^{\infty}(L_s^2(\rz^{dp}))$  the sequence $(\Tr[\varrho^{(p)}_{n} A])_{n\in\nz^*}$ converges. Then there exists a unique Borel probability measure $\mu_0$ on $L^2(\rz^d)$ invariant with respect to the unitary group $U(1)$ and such that for any $p\in\nz^*$ and any  $A\in\L^{\infty}(L_s^2(\rz^{dp}))$,
\bean
\lim_{n\to\infty} \Tr[\varrho^{(p)}_{n} A] =\Tr[\varrho_{\infty}^{(p)} A]\,, \quad \mbox{ with }  \quad
\varrho_{\infty}^{(p)}=\int_{L^{2}(\rz^d)} |z^{\otimes p}\rangle\langle z^{\otimes p}| \, d\mu_0(z)\,.
\eean
Moreover, the measure $\mu_0$ is  concentrated on the unit ball $B_{}(0,1)$ of $L^2(\rz^d)$ centered at the origin and of radius one (i.e.: $\mu_0(B_{}(0,1))=1)$.
\end{prop}
Actually, the measure $\mu_0$ is the unique Wigner measure of the sequence  $(\varrho_n)_{n\in\nz^*}$ (see Subsection \ref{sub.wigner} for definition and details).
Once this is understood we can consider the problem of rate of convergence for more general correlated states.

\begin{thm}
\label{main-th}
Let $(\alpha(n))_{n\in\nz^*}$ be a sequence of positive numbers with $\lim\alpha(n)=\infty$ and such that $(\frac{\alpha(n)}{n})_{n\in\nz^*}$ is bounded.
Let $(\varrho_{n})_{n\in\nz^*}$ and $(\varrho_{\infty}^{(p)})_{p\in\nz^*}$ be two sequences of density matrices with $\varrho_n\in\L^1(L^2_s(\rz^{dn}))$ and $\varrho_\infty^{(p)}\in\L^1(L^2_s(\rz^{dp}))$  for each $n,p\in\nz^*$.  Assume that there exist $C_0>0$, $C>2$ and $\gamma\geq 1$ such that for all
$n,p\in\nz^*$ with $n\geq \gamma p$:
\bea
\label{init-ineq}
\left\|\varrho^{(p)}_{n}-\varrho_{\infty}^{(p)}\right\|_1\leq
 C_0 \frac{C^p}{\alpha(n)}\,.
\eea
 Then for any $T>0$ there exists $C_T>0$ such that for all $t\in[-T,T]$ and all $n,p\in\mathbb{N}^*$ with $n\geq \gamma p$,
\bea
\label{main-ineq}
\left\|\varrho^{(p)}_{n}(t)-\varrho_{\infty}^{(p)}(t)\right\|_1\leq
 C_T \frac{C^p}{\alpha(n)}\,,
\eea
where
\bean
\varrho_{\infty}^{(p)}(t)=\int_{L^{2}(\rz^d)} |z^{\otimes p}\rangle\langle z^{\otimes p}| \, d\mu_t(z)\,,
\eean
with $\mu_t=(\Phi_t)_\sharp\mu_0$ is the push-forward of the initial measure $\mu_0$ (given in Proposition \ref{definetti}) by  the well defined and continuous Hartree  flow $\Phi_t$ on $L^2(\rz^d)$  of the equation \eqref{hartree} (given in Subsection \ref{sec.classical-quantum}).
\end{thm}
\begin{remarks}\ \,
\begin{itemize}
\item [1)] Our result holds true in a more general framework. We can replace
$L^2(\rz^d)$ by any separable Hilbert space $\Z$,  $-\Delta$  by any self-adjoint operator $h_0$, and $V$ by any two-particle bounded interaction (see Subsection \ref{sec.classical-quantum}). So from now on we will consider this setting, which has the advantage of covering several situations: e.g. either finite or infinite dimensional systems, as well as semi or non relativistic ones.
\item [2)] The assumption \eqref{init-ineq} implies that we can apply Proposition \ref{definetti} and hence obtain the existence of the initial measure $\mu_0$ at $t=0$.
\item [3)] The condition $C>2$ in the main assumption of Theorem \ref{main-th} can be replaced by $C>0$ at the cost of slightly changing the conclusion, by replacing $C$ in \eqref{main-ineq}  by $C+2$.
\item [4)] We can apply Theorem \ref{main-th} backward in time. So, if the estimates \eqref{main-ineq} hold  true at a given time $t$,  then \eqref{init-ineq}  should also hold at time $t=0$. This answers the question of improvement of the rate of convergence under the action of the quantum evolution.   Indeed, if we suppose that inequalities  \eqref{main-ineq} hold with a faster rate of convergence $\beta(n)$, $\lim \frac{\alpha(n)}{\beta(n)}=0$, then the ``initial'' estimate \eqref{init-ineq} should also hold with $\beta(n)$ instead of $\alpha(n)$ by backward evolution.
\item[5)] The proof of Theorem \ref{main-th} allows to start with a rate of convergence $\alpha(n)$ faster than $1/n$ at time $t=0$. However,  we can't recover a better convergence at time $t\neq 0$. This is why we have restricted $\alpha(n)$ to be of order $n$ or less.   However, this feature do not seem to be an artefact of the proof: numerical simulations on product states indicate a $1/n$ order of convergence  even when at time $t=0$ the reduced density matrices coincide
with their limit.
\end{itemize}
\end{remarks}

The mathematical analysis  of the mean field limit is quite rich and indeed there are several approaches and techniques applicable to this problem. For example coherent states analysis \cite{GiVe1,GiVe2,Hep}, BBGKY hierarchy method \cite{Spo}, Egorov type theorem \cite{FGS,FKP,FKS}, Wigner measures approach \cite{AmNi1,AmNi4,QB} or deviation estimates \cite{KP,Pi}. Hence the combination of these different techniques may lead to interesting results. The proof of our main Theorem \ref{main-th} relies on two ingredients: an Egorov type theorem proved
in \cite{FGS,FKP} and a Wigner measures characterization of the limit points of reduced density matrices studied in \cite{AmNi1,AmNi2,AmNi3}. So the first step is to use second quantization formalism and Wick observables, then the result in \cite{FGS,FKP} provides the asymptotics of time-evolved Wick observables as
\bea
\label{int-expa}
e^{it \mathbf{H}_n} \,\,b^{Wick} e^{-it \mathbf{H}_n} \; _{|L^2_s(\mathbb{R}^{dn})}=b(t)^{Wick} \;_{|L^2_s(\mathbb{R}^{dn})}+R(n),
\eea
where $\lim_{n\to\infty}R(n)=0$ in some specific sense and where $b(t)^{Wick}$ is an infinite sum
of Wick operators  with time-dependent kernels or symbols (see Subsections \ref{sub.wick} and \ref{sub.expansion}). The mean field expansion \eqref{int-expa} gives actually the convergence of the correlation functions (\ref{eq.1}). So that, if we use the idea of  Wigner measures extended to this framework in \cite{AmNi1}, we can obtain the rate of convergence for the quantities (\ref{eq.1}). Once this is proved, one can get the announced   trace norm  estimates for the difference between reduced density matrices.

The article is organized as follows. The second quantization formalism and Wick symbolic calculs is recalled in Subsection \ref{sub.wick}. The mean field expansion is explained in
Subsection \ref{sub.expansion} while the the  quantum and classical dynamics are introduced in Subsection \ref{sec.classical-quantum}. In Section \ref{sec:RDM}, we analyse the relationship between reduced density matrices (RDM) and Wigner measures and provide the proof of Proposition \ref{definetti}. Our main result is proved in Section \ref{sec.rate} with some preliminary lemmas. Examples and numerical simulations are discussed in the last Section \ref{sec.ENS}.

\section{Mean field expansion}
\label{sec.MFE}
The mean field theory is  concerned with quantum dynamical systems which preserve the number of particles and can be worked out in the setting of multi-particles Schr\"odinger operators \eqref{schrod}. Nevertheless, it is advantageous  to use the more general setting of second quantization, as the reader will notice throughout the following sections.
Actually, the Hamiltonian \eqref{schrod} can be reformulated as
$$
\mathbf{H}_n=\varepsilon^{-1}H_{\varepsilon_{|L^2_s(\mathbb{R}^{dn})}},\,
$$
with $\varepsilon=\frac{1}{n}$ and $H_{\varepsilon}$ a Hamiltonian on the symmetric Fock space over $L^2(\mathbb{R}^d)$ given by
\begin{eqnarray}
\label{ham_fock}
  && H_{\varepsilon}=\varepsilon \int_{\rz^{d}}
\nabla a^{*}(x)\nabla
a(x)~dx+\frac{\varepsilon^2}{2} \int_{\rz^{2d}}V(x-y)a^{*}(x)a^{*}(y)a(x)a(y)~dxdy\,,
\end{eqnarray}
where $a,a^*$ are the usual creation-annihilation operator-valued distributions, i.e.:
$$
[a(x),a^*(y)]=\delta(x-y)\,,\;\; [a^*(x),a^*(y)]=0=[a(x),a(y)]\,.
$$
Our investigation of the mean field approximation for the quantum dynamics
\eqref{schrod} is made through the analysis of the Hamiltonian \eqref{ham_fock}. The strategy relies on a specific Schwinger-Dyson expansion of the time dependent correlation functions \eqref{correl} elaborated in
\cite{FGS,FKP} combined to  some tools (Wigner measures) from semiclassical analysis extended to infinite dimensional setting in \cite{AmNi1}. The Schwinger-Dyson expansion, called here mean field expansion, is explained in Subsection \ref{sub.expansion} and leads naturally  to the consideration of several multiple commutators which we need to normal order using Wick's theorem. So, for reader convenience we recall some basic results on normal ordering and Wick operators written in more systematic and in some sense more efficient way: it makes possible the use of a symbolic calculus, for an algebra of Wick operators, similar to the pseudo-differential calculus in finite dimension
(for other presentations of second quantization see \cite{Ber,DG}).

\subsection{Wick calculus}
\label{sub.wick}
From now on we will wok in a general setting. Let $\mathscr{Z}$ be a separable Hilbert space. The symmetric Fock space over $\mathscr{Z}$ is the Hilbert space
\begin{eqnarray*}
\mathscr{H}=\underset{n=0}{\overset{\infty}{\oplus}}\vee^n
\mathscr{Z}\,,
\end{eqnarray*}
where $\vee^n \mathscr{Z}$ denotes the $n$-fold symmetric tensor product.
The orthogonal projection of $\Z^{\otimes n}$ onto the closed subspace $\vee^n\Z$ is denoted by $\S_{n}$. The dense subspace of finite particle vectors is denoted by
$$
\H_0=\underset{n\geq 0}{\overset{alg}{\oplus}}\vee^n \Z\,.
$$
So, the creation and annihilation operators  $a^*(f)$ and $a(f)$, parameterized by $\hbarr>0$, are then defined by :
\begin{eqnarray*}
a(f) \varphi^{\otimes n}&=&\sqrt{\hbarr n} \; \;\la f,\varphi\ra \varphi^{\otimes (n-1)}\\
a^*(f)\varphi^{\otimes n}&=&\sqrt{\hbarr (n+1)}  \;\;\S_{n+1}
(\;f\otimes \varphi^{\otimes n})\,,\;\;\, \forall \varphi\in\Z.
\end{eqnarray*}
They extend to closed operators, they are adjoint and satisfy the canonical commutation relations (CCR):
\begin{eqnarray*}
%\label{ccr}
[a(f),a^*(g)]=\hbarr\la f,g\ra\, 1, \;\;\;[a^*(f),a^*(g)]=[a(f),a(g)]=0\,,\quad\forall f,g\in \Z.
\end{eqnarray*}
The Weyl operators are
$$
W(f)=e^{\frac{i}{\sqrt{2}} [a^*(f)+a(f)]}\,,\quad f\in\Z\,,
$$
and they satisfy the  Weyl commutation relations
\begin{eqnarray*}
%\label{eq.Weylcomm}
W(f) W(g)=e^{-\frac{i\hbarr}{2} {\rm Im}(f,g)} \;W(f+g ), \quad\forall f,g\in\Z\,.
\end{eqnarray*}
For any (possibly unbounded) operator $A:\D(A)\subset\Z \to \Z,$
we define $\d\Gamma(A)$ as
\begin{eqnarray*}
 \d\Gamma(A)_{|\vee^{n,\textrm{alg}}\D(A)}=\hbarr
\sum_{k=1}^{n}1^{\otimes (k-1)}\otimes A\otimes 1^{\otimes (n-k)}\,,
\end{eqnarray*}
where $\vee^{n,\textrm{alg}}\D(A)$ denotes the $n$-fold algebraic symmetric tensor product of $\D(A)$. \\
Any Wick operator preserving the number of particles could be written in the case of $\Z=L^2(\rz^d)$ as a quadratic form using the integral formula
\bean
b^{Wick}=\varepsilon^{k} \int_{\mathbb{R}^{2kd}} \prod_{i=1}^k a^*(x_i) B(x_1,\cdots,x_k;y_1,\cdots,y_k) \prod_{j=1}^k a(y_j)\;
 dx_1\cdots dx_k dy_1\cdots dy_k,
\eean
with $B(x_1,\cdots,x_k;y_1,\cdots,y_k)$ denotes the distribution kernel of the operator $B$ on $L^{2}(\mathbb{R}^{kd})$.
For general Hilbert spaces, this formula can be generalized as follows.
\begin{defn}[Class of symbols]
\label{def.symbol}
For any $p,q\in \mathbb{N}$, define $\P_{p,q}$ to be the space
 of homogeneous complex-valued polynomials on $\Z$  such that $b\in\P_{p,q}$
 if and only if there exists a (unique) bounded operator $\tilde b\in\L(\vee^p\Z,\vee^q\Z)$ such that for all $z\in\Z$:
\bea
\label{eq.id}
b(z)=\la z^{\otimes q}, \tilde{b} \,z^{\otimes p}\ra\,.
\eea
\end{defn}
We will often use the identification between homogeneous polynomials $b\in\P_{p,q}$
and their associated operators $\tilde b\in\L(\vee^p\Z,\vee^q\Z)$ according to \eqref{eq.id}.  The algebraic sum
$$
\P=\underset{p,q\geq 0}{\overset{alg}{\oplus}} \P_{p,q}
$$
is clearly an algebra of polynomials.
 These spaces $\P$ and $\P_{p,q}$ play a role similar, in some sense, to classes of symbols in pseudo-differential calculus. For this reason we sometimes call the polynomials $b\in\P$ symbols (see for instance \cite{BoLe}). The subspace of $\P_{p,q}$ made of polynomials $b$ such that
$\tilde{b}$ is a compact operator  is
denoted by $\mathscr{P}^{\infty}_{p,q}$ and
$$
\P^\infty=\underset{p,q\geq 0}{\overset{alg}{\oplus}} \mathscr{P}^{\infty}_{p,q}.
$$
\begin{defn}[Wick operators]
A {\it Wick operator} with symbol $b\in \P_{p,q}$ is a linear operator  $b^{Wick}$ with domain $\H_0$
defined as
\begin{eqnarray*}
%\label{wick}
b^{Wick} \,_{|\vee^n \Z}=1_{[p,+\infty)}(n)\frac{\sqrt{n!
(n+q-p)!}}{(n-p)!} \;\hbarr^{\frac{p+q}{2}} \;\S_{n-p+q}\left(\tilde{b}\otimes 1^{\otimes(n-p)}\right)\,,
\end{eqnarray*}
where $\tilde{b}$ denotes the operator associated to the symbol $b$ according to
\eqref{eq.id}.
\end{defn}
Remark that for simplicity we have used the notation $b^{Wick}$ without stressing the dependence on the  scaling parameter $\hbarr$.

An interesting feature of the above quantization is that it maps the algebra of symbols or polynomials $\P$ into an algebra of operators in the Fock space. In particular, the composition of two given Wick operators $b_1^{Wick}$ and $b_2^{Wick}$ is again a Wick operator $c^{Wick}$. More interesting is that its symbol $c$ belongs to $\P$ and  is given by an explicit formula like in  pseudo-differential calculus of finite dimension. This comparison goes much further in fact, and the commutator of Wick operators is a sum of quantized  multiple Poisson brackets with $\hbarr$ playing the role of a semi-classical parameter.

Let us introduce the precise meaning of the multiple Poisson brackets. Remark that all polynomials in $\P_{p,q}$ admit Fr\'echet differentials and therefore they all have directional derivatives. Remark also that we don't need a particular conjugation on  the Hilbert space $\Z$ in order to define the derivatives $\partial_{\bar z}$ and $\partial_z$. In fact, for $b\in\P_{p,q}$ we define
\begin{eqnarray*}
 \partial_{\overline{z}} b(z)[u]=\bar\partial_r b(z+ r u)_{|r=0} ,&& \partial_{z} b(z)[u]=
 \partial_{r} b(z+ r u)_{|r=0}\,,
\end{eqnarray*}
where $\bar\partial_r, \partial_r$ are the usual derivatives over
$\mathbb{C}$. Moreover, $\partial_{z}^{k}b(z)$ naturally belongs to
$(\vee^{k}\Z)^{*}$ (i.e.: $k$-linear symmetric functionals) while
$ \partial_{\overline{z}}^{j}b(z)$ is identified via the scalar
product with an element of
$\vee^{j}\Z$, for any fixed $z\in \Z$. For $b_{i}\in \P_{p_{i},q_{i}}$,
$i=1,2$ and $k\in\mathbb{N}$, set
\begin{equation*}
\partial_z^k b_1  \cdot\partial_{\bar z}^k b_2 (z) =\la \partial_z^k b_1(z),
\partial_{\bar z}^k b_2(z)\ra_{(\vee^k \Z)^{*},\vee^{k}\Z}=\partial_z^k b_1(z)[\partial_{\bar z}^k b_2(z)]\; \in\P_{p_1+p_2-k, q_1+q_2-k}\quad.
\end{equation*}
The multiple {\it Poisson brackets} are defined by
\begin{eqnarray}
\label{multi-brack}
\{b_1,b_2\}^{(k)}=\partial^k_z b_1
\cdot\partial^k_{\bar z} b_2 -\; \partial^k_z b_2 \cdot\partial^k_{\bar z} b_1 \quad \mbox{
and } \quad
\{b_1,b_2\}=\{b_1,b_2\}^{(1)}.
\end{eqnarray}

\begin{prop}
\label{wick-calc}
Let $b_{1} \in \mathscr P_{p_{1},q_{1}}$ et $b_{2} \in \mathscr P_{p_{2},q_{2}}$. For all $k \in \{ 0,..., \min{(p_{1},q_{2})} \}$, the polynomial $\partial_{z}^{k}b_{1}\cdot \,\partial_{\bar z}^{k}b_{2}$ belongs to $\mathscr P_{p_{1}+p_{2}-k,q_{1}+q_{2}-k}$ with the following formulas holding true on $\H_0$:
\begin{equation*}
b_{1}^{Wick} \circ b_{2}^{Wick} =\Big[ \sum_{k=0}^{\min(p_{1},q_{2})}\frac{\varepsilon^{k}}{k!} \partial_{z}^{k}b_{1}\cdot\partial_{\bar z}^{k}b_{2} \Big]^{Wick}.
\end{equation*}
\begin{equation*}
[b_{1}^{Wick},b_{2}^{Wick}]=\sum_{k=1}^{\max(\min(p_{1},q_{2}),\min{(p_{2},q_{1})})}\frac{\varepsilon^{k}}{k!} \Big[\{b_{1},b_{2} \}^{(k)} \Big]^{Wick}.
\end{equation*}
\end{prop}

\subsection{Classical and Quantum dynamics}
\label{sec.classical-quantum}
Consider a polynomial $Q\in\P_{2,2}$ such that $\tilde{Q}\in\L(\vee^2\Z)$ is bounded and symmetric.
In all the sequel we consider  the many-body quantum Hamiltonian of bosons to be the operator defined by
\begin{eqnarray}
\label{hamiltonian}
H_\hbarr=\d\Gamma(\tilde h_0)+ Q^{Wick},
\end{eqnarray}
where $\tilde h_0$ is  a given self-adjoint operator on $\Z$ with domain $\D(\tilde h_0)$. By standard perturbation theory, and thanks to the conservation of the number of particles, it is easy to prove that $H_\hbarr$ is essentially self-adjoint on
$\D(\d\Gamma(\tilde h_0))\cap\H_0$. We denote respectively the time evolution of the perturbed and the free quantum system by
\bean
U(t)=e^{-i\frac{t}{\hbarr} H_{\hbarr}}
  \quad \mbox{ and } \quad U_0(t)=e^{-i\frac{t}{\hbarr} \d\Gamma(\tilde h_0)}\,.
\eean

\bigskip
 It is known that in the mean field limit we obtain the Hartree equation \eqref{hartree}, when the many-body Schr\"odinger Hamiltonian \eqref{schrod} is considered. In our abstract setting the limit equation has the energy functional
\begin{equation*}
    \label{eq.enfunct}
    h(z)=\la z,\tilde h_0 z\ra+ Q(z)\,,\;\;\;z\in\D(\tilde h_0)\,,
\end{equation*}
which is actually the Wick symbol of the quantum Hamiltonian  \eqref{hamiltonian}. So, the associated  nonlinear field equation reads
\begin{eqnarray}
\label{hartree-abs}
i \partial_t z_t=X(z_t)
\end{eqnarray}
with $X:\D(\tilde h_0)\to \Z$ is the vector field given by $ X(z)=\tilde h_0 z+\partial_{\bar z} Q(z)$. In order to solve this equation we write it in the integral form
\begin{eqnarray}
\label{hartree.int}
 z_t=e^{-i t \tilde h_0} z_0-i\int_0^t e^{-i (t-s) \tilde h_0} \,\; \partial_{\bar z}Q(z_s) \, ds, \;\mbox{ for } \;z_0\in\Z\,.
\end{eqnarray}
Since $\tilde Q$ is a bounded operator then a standard fixed point argument implies
 that (\ref{hartree.int}) admits a unique continuous local solution for each initial condition $z_0\in\Z$. Thanks to the conservation of the Hilbert norm on $\Z$ we see that any local solution extends to a global continuous one. Therefore, we have a well defined global continuous flow on $\Z$ which we denote  by
$\Phi:\rz\times \Z\to \Z$. In other terms  $\Phi$ is a $C^0$-map satisfying $\Phi_{t+s}(z)=
\Phi_{t}\circ\Phi_{s}(z)$ and $z_t:=\Phi_t(z_0)$ solves (\ref{hartree.int}) for any $z_0\in\Z$.
 Moreover, if $\rz\ni t\mapsto z_t$  is the solution of $(\ref{hartree.int})$
 and $Q_{t}$ is the polynomial in $\P_{2,2}$ given as $Q_{t}(z)=Q(e^{-it\tilde h_0}z)$, then the
 curve $w_t=e^{i t\tilde h_0} z_t$ solves the differential equation
$$
\frac{d}{dt}\, w_t=-i \partial_{\bar z} Q_t(w_t)\,.
$$
Hence, a simple computation yields for any $b\in\P_{p,q}$ the identity
\bean
\frac{d}{dt}\, b(w_t)&=& i\partial_{ z}Q_t(w_t) [\partial_{\bar z} b(w_t)]+
\partial_{z} b(w_t)[-i \partial_{\bar z}Q_t(w_t)]\\&=&i \{Q_t,b\}(w_t),
\eean
where the brackets are defined according to \eqref{multi-brack}. So, we obtain the  following Duhamel formula for all $t\in\rz$:
\bea
\label{class-integ-form}
b(z_t)=b_t(w_0)+i\int_{0}^{t}\left\{Q_{t_1},b_t\right\}(w_{t_1})~dt_{1}\,,
\eea
with $t\in\rz\mapsto z_t$ a (mild) solution of the nonlinear field equation \eqref{hartree-abs} and
$w_t=e^{i t\tilde h_0} z_t$.

\subsection{Mean field expansion}
\label{sub.expansion}
The main point is to study the time evolution of Wick operators with respect to the small mean field parameter $\hbarr$ which is  essentially the inverse of the number of particles.  This was done in \cite{FGS, FKP} and in fact we can prove in some sense that
\bea
\label{formal_wick}
e^{i\frac{t}{\varepsilon} H_\varepsilon} \,\,A^{Wick} e^{-i\frac{t}{\varepsilon} H_\varepsilon}=A(t)^{Wick}+R(\varepsilon),
\eea
with $R(\varepsilon)\to 0$ when $\varepsilon\to 0$ (see \cite{FGS, FKP} and also \cite[Thm.~5.5]{AmNi1}) and where $A(t)^{Wick}$ is an infinite sum of Wick operators with time-dependent symbols related to the Hartree dynamics.  In order to prove \eqref{formal_wick}, we use an iterated integral formula (the so-called Dyson-Schwinger expansion) with a specific use of Wick calculus (Proposition \ref{wick-calc}) in order to expand commutators of Wick operators with respect to the $\hbarr$ parameter.
We will work in the interaction representation. Hence, the following notation is useful
\begin{equation*}
%\label{eq.wickfree}
b_{t}=b\circ e^{-it\tilde h_0}~:~\Z\ni z \mapsto b_{t}(z)=b(e^{-it\tilde h_0}z)\,\,,
\end{equation*}
for  $b\in\P$ and  $t\in \rz$ (remark that $ b_{t}$ belongs to $\P$). We also know  that  multiple commutators in the Schwinger-Dyson expansion  lead to Wick operators with  multiple Poisson brackets symbols. For this reason  we make the following definition.

\begin{defn}
For $m\in \nz$ and $(t_1,\cdots,t_m,t)\in\mathbb{R}^{m+1}$, we associate to any
$b\in \P_{p,p}$ the polynomial:
\begin{eqnarray}
\label{cnr}
C^{(0)}_0(t)=b_t \quad \mbox{ and } \quad
C^{(m)}_0(t_m,\cdots,t_1,t)=\;  \bigg\{Q_{t_m}, \cdots, \bigg\{Q_{t_1},b_t
\bigg\}\cdots\bigg\}
\in\P_{p+m,p+m}\,.
\end{eqnarray}
For simplicity the dependence of $C_0^{(m)}(t_m,\cdots,t_1,t)$ on the symbol $b$
is not made explicit and sometimes we will  write $C_0^{(m)}$ for shortness.
\end{defn}
The above polynomials $C^{(m)}_0$ satisfy the following iteration formula.
\begin{lem} For $m\in\nz$ and $(t_1,\cdots,t_m,t)\in\mathbb{R}^{m+1}$,
\label{iter-com}
\bean
\frac{1}{\hbarr} \big[Q_{t_m}^{Wick}, C^{(m-1)}_0(t_{m-1},\cdots,t_1,t)^{Wick}\big]&=& C^{(m)}_0(t_{m},\cdots,t_1,t)^{Wick} \\&&+
\frac{\hbarr}{2} \left(\bigg\{Q_{t_m}, C^{(m-1)}_0(t_{m-1},\cdots,t_1,t)\bigg\}^{(2)}\right)^{Wick}\,.
\eean
\end{lem}
\proof This is a straightforward consequence of the definition  of $C^{(m)}_0$ and the composition formula in Proposition
\ref{wick-calc}. \fin

\bigskip
\noindent
We consider a sequence $(\varrho_n)_{n\in\nz^*}$ of density matrices such that $\varrho_n\in \L^1(\vee^n\Z)$. For shortness, we denote
$$
\varrho_{{n}} (t)=U(t)\;\varrho_{{n}}\;U(t)^* \quad \mbox{ and } \quad \tilde\varrho_{{n}} (t)= U_0(t)^*\;\varrho_{{n}}(t)\;U_0(t)  \quad\mbox{ with }\quad \hbarr=\frac{1}{n}\,,
$$
and for simplicity write $A^{Wick}$ for the Wick operator with symbol $\la z^{\otimes p}, Az^{\otimes p}\ra$ with $A\in\L(\vee^p\Z)$.
\begin{prop}
\label{prop-1}
Let $(\varrho_n)_{n\in\nz^*}$ be a sequence of density matrices such that $\varrho_n\in \L^1(\vee^n\Z)$ for each $n\in\nz^*$. Then for any $n,p\in\nz^*$ such that $p\leq n$, $A\in\L(\vee^p\Z)$, $M\in\nz^*$ and $ t\in\rz$:
\bea
\label{formula-m}
{\rm Tr}[\varrho_{{n}} (t) \; A^{Wick}] &=&\sum_{k=0}^{M-1}  i^k
 \int_0^t dt_1\cdots\int_0^{t_{k-1}} dt_k \;{\rm Tr}\left[\varrho_{n}\;
C_0^{(k)}(t_k,\cdots,t_1,t)^{Wick}\right] \nonumber \\ \nonumber
 + \frac{\varepsilon}{2}
  \sum_{k=1}^{M} i^{k}&&\hspace{-.3in}
\int_0^t dt_1\cdots\int_0^{t_{k-1}} dt_k\;
\Tr\left[\tilde\varrho_n(t_k)   \left(\bigg\{Q_{t_{k}},C^{(k-1)}_{0}(t_{k-1},\cdots,t_1,t)
\bigg\}^{(2)}\right)^{Wick} \right]\\
+i^{M} &&\hspace{-.3in}\int_0^t
dt_1\cdots\int_0^{t_{M-1}} dt_M \;\Tr\left[ \tilde\varrho_n(t_M) \;C^{(M)}_0(t_M,\cdots,t_1,t)^{Wick}\right]\,,
\eea
with $C_0^{(k)}$ given in \eqref{cnr} and the multiple Poisson bracket defined in \eqref{multi-brack}.
\end{prop}
\proof
The expansion is obtained by iteration. Let $b\in\P_{p,p}$ then
$$
\frac{d}{dt} \,U(t)^* U_0(t) b^{Wick} U_0(t)^* U(t)_{|\vee^n\Z}=\frac{i}{\hbarr}
U(t)^* U_0(t) [Q_t^{Wick}, b^{Wick}] U_0(t)^* U(t)_{|\vee^n\Z}\,.
$$
A simple integration yields
\bea
\label{eq.iter.dh}
U(t)^* U_0(t) b^{Wick} U_0(t)^* U(t)_{|\vee^n\Z}= b^{Wick} \,_{|\vee^n\Z}+\frac{i}{\hbarr} \int_0^t
U(t)^* U_0(t) [Q_t^{Wick}, b^{Wick}] U_0(t)^* U(t)_{|\vee^n\Z}\,.
\eea
Taking $b^{Wick} = U_0(t)^* A^{Wick} U_0(t)=A_t^{Wick}$ in the above formula, gives
\bean
U(t)^* A^{Wick}U (t)_{|\vee^n\Z} &=&U_0(t)^* A^{Wick}U_0(t)_{|\vee^n\Z}\\ &&+ \frac{i}{\hbarr}
 \int_0^t dt_1 U(t_1)^*
U_0(t_1) \left[Q_{t_1},A_t^{Wick}\right] U_0(t_1)^*
U(t_1)_{|\vee^n\Z} \,.
\eean
Hence using  Lemma \ref{iter-com}, we get
\bean
U(t)^* A^{Wick}U (t)_{|\vee^n\Z} &=&U_0(t)^* A^{Wick}U_0(t)_{|\vee^n\Z}\\ &&+ i
 \int_0^t dt_1 \, \underbrace{U(t_1)^*
U_0(t_1) \,C_0^{1}(t_1,t)^{Wick} \, U_0(t_1)^*
U(t_1)}_{(\mathrm{T})} \,_{|\vee^n\Z}
\\ &&+i \frac{\hbarr}{2}
 \int_0^t dt_1 U(t_1)^*
U_0(t_1) \left(\bigg\{Q_{t_1},A_t\bigg\}^{(2)}\right)^{Wick} U_0(t_1)^*
U(t_1)_{|\vee^n\Z} \,.
\eean
Remark that the first two terms in the right hand side are of order $O(1)$ while the last one is of order $O(\hbarr)$.
By using again \eqref{eq.iter.dh} to expand the term (T) above with $b=C^{1}_0(t_1,t)$ we obtain, after taking the trace with $\varrho_n$, the formula   \eqref{formula-m} for $M=2$.
So, iterating this process  $M-1$ times and following the same scheme of splitting commutators into two parts one of order $O(1)$ and the second of order $O(\hbarr)$, we get
\bean
U(t)^* A^{Wick}U (t)_{|\vee^n\Z} &=& \\ &&\hspace{-1.1in}\sum_{k=0}^{M-1}
 i^{k} \int_0^t
dt_1\cdots\int_0^{t_{k-1}} dt_k \;
C^{(k)}_0(t_k,\cdots,t_1,t)^{Wick}\;\\ && \hspace{-1.1in} + i^{M} \int_0^t
dt_1\cdots\int_0^{t_{M-1}} dt_M \; U(t_M)^*
U_0(t_M)\;
\left[C^{(M)}_0(t_M,\cdots,t_1,t)\right]^{Wick} \;U_0(t_M)^*
U(t_M) \\ &&\hspace{-1.5in}+\frac{\hbarr}{2} \sum_{k=1}^{M-1}
 i^{k} \int_0^t
dt_1\cdots\int_0^{t_{k-1}} dt_k \; U(t_k)^*
U_0(t_k) \left( \bigg\{Q_{t_k};
C^{(k-1)}_0(t_{k-1},\cdots,t_1,t)\bigg\}^{(2)}\right)^{Wick} U_0(t_k)^*
U(t_k)\,.
\eean
Hence, by taking the trace with $\varrho_n$ we prove the proposition. \fin

The next step is to let  $M\to\infty$ in the formula \eqref{formula-m}. But to do this we need to prove some estimates which guarantee the absolute convergence of these series.
\begin{lem}
\label{tech.lem.1}
For any  $b\in\P_{p,p}$ the symbols $\{Q_s,b_t\}^{(2)}\in\P_{p,p}$ and $C^{(m)}_0\in\P_{p+m,p+m}$ with the following inequalities holding true:\\
(i)
\begin{eqnarray*}
\left\|\widetilde{\{Q_s,b_t\}^{(2)}}\right\|_{\mathscr{L}(\vee^{p}\Z)}
\leq  \; 4p(p-1) \;\|\tilde Q\| \;\|\tilde b\|_{\mathscr{L}(
\vee^p\Z)}\,.
\end{eqnarray*}
(ii) For any $m\in \nz$,
\begin{eqnarray*}
&&\hspace{-.5in}\left\|\widetilde{C^{(m)}_0}(t_m,\cdots,t_1,t)\right\|_{\mathscr{L}(\vee^{p+m}\Z)}\leq 4^{m} \;\ds \;\frac{(p+m-1)!}{(p-1)!} \;\|\tilde Q\|^m
\;\|\tilde b\|_{\mathscr{L}(\vee^p\Z)}\,.
\end{eqnarray*}
Here $\widetilde{\{Q_s,b_t\}^{(2)}}$ and $\widetilde{C^{(m)}_0}$ are respectively the operators associated to the polynomials $\{Q_s,b_t\}^{(2)}$ and $C^{(m)}_0$ according to Definition \ref{def.symbol}.
\end{lem}
\proof
See \cite[Lemma 5.8, 5.9]{AmNi1}.
\fin
\begin{prop}
\label{prop-2}
Let $(\varrho_n)_{n\in\nz^*}$ be a sequence of density matrices such that $\varrho_n\in \L^1(\vee^n\Z)$ for each $n\in\nz^*$. Then for any $n,p\in\nz^*$ such that $p\leq n$, $A\in\L(\vee^p\Z)$ and $ |t|<\frac{1}{8||\tilde Q||}$ and $\hbarr=\frac{1}{n}$:
\bea
\label{form-inf}
{\rm Tr}[\varrho_{{n}} (t) \; A^{Wick}] &=&\sum_{k=0}^{\infty}  i^k
 \int_0^t dt_1\cdots\int_0^{t_{k-1}} dt_k \;{\rm Tr}\left[\varrho_{n}\;
C_0^{(k)}(t_k,\cdots,t_1,t)^{Wick}\right] \nonumber \\ \nonumber
 + \frac{\varepsilon}{2}
  \sum_{k=1}^{\infty} i^{k}&&\hspace{-.3in}
\int_0^t dt_1\cdots\int_0^{t_{k-1}} dt_k\;
\Tr\left[\tilde\varrho_n(t_k)   \left(\bigg\{Q_{t_{k}},C^{(k-1)}_{0}(t_{k-1},\cdots,t_1,t)
\bigg\}^{(2)}\right)^{Wick} \right]\,,
\eea
with  $C_0^{(k)}$ are given in \eqref{cnr}, the multiple Poisson bracket defined in \eqref{multi-brack} .
\end{prop}
\proof
Proposition \ref{prop-1} says that
$$
{\rm Tr}[\varrho_{{n}} (t) \; A^{Wick}] =\sum_{k=0}^{M-1} C_k + \frac{\hbarr}{2} \sum_{k=1}^{M} B_k + R_M\,,
$$
where $A_k,B_k$ and $R_M$ are short notations for the terms appearing in \eqref{formula-m}.
Applying Lemma \ref{tech.lem.1} and using the fact that we integrate $k$-times, we get for $n\geq p+k$
\begin{eqnarray}
\big| C_k\big|&\leq& 4^{k} \frac{(p+k-1)!}{(p-1)! k!} (|t| \,||\tilde Q||)^k  \, ||A||\,,  \nonumber\\ 
\big| B_k\big|&\leq &4^{k} (p+k-1) (p+k-2) \frac{(p+k-1)!}{(p-1)! k!} (|t| \,||\tilde Q||)^k  \, ||A||\,,  \\
\big| R_M\big|&\leq& 4^{M} \frac{(p+M-1)!}{(p-1)! M!} (|t| \,||\tilde Q||)^M  \, ||A||\,.  \nonumber\label{eq.est-err}
\end{eqnarray}
Therefore, using the bound $C^k_{p+k-1}\leq 2^{p+k-1}$, we see that for times $ |t|<\frac{1}{8||\tilde Q||}$ the two series
$\displaystyle\sum_{k=0}^{M-1} C_k$ and $\displaystyle\sum_{k=1}^{M} B_k$  are absolutely convergent and  $R_M\to 0$ when
$M\to\infty$. \fin

\begin{prop}
\label{prop-3}
Let $(\varrho_n)_{n\in\nz^*}$ be a sequence of density matrices such that $\varrho_n\in \L^1(\vee^n\Z)$ for each $n\in\nz^*$. Then for any $C>2$ there exists $C_0>0$ such that  for any $n,p\in\nz^*$, $p\leq n$, $A\in\L(\vee^p\Z)$, $ |t|<\frac{1}{16||\tilde Q||}$ and $\hbarr=\frac{1}{n}$:
\bea
\label{est-inf}
\left|{\rm Tr}[\varrho_{{n}} (t) \; A^{Wick}] -\sum_{k=0}^{\infty}  i^k
 \int_0^t dt_1\cdots\int_0^{t_{k-1}} dt_k \;{\rm Tr}\left[\varrho_{n}\;
C_0^{(k)}(t_k,\cdots,t_1,t)^{Wick}\right]\right| \leq  C_0\,\frac{C^{p}}{n} ||A|| \,.
\eea
\end{prop}
\proof
This follows by Proposition \ref{prop-2} and estimate \eqref{eq.est-err}.
In fact, we see that the left hand side of \eqref{est-inf} is bounded by
\bean
\frac{2^{p-1}}{2n} \sum_{k=1}^\infty (k+p)^2\big(8|t|\,||\tilde Q||\big)^k \,||A||\leq\frac{2^{p}}{2n} \sum_{k=1}^\infty (k^2+p^2)\big(8|t|\,||\tilde Q||\big)^k \,||A||\leq
\frac{2^p}{n} (3+p^2) \,||A|| \,.
\eean
Taking $C_0=\ds\max_{p\geq 1} \frac{2^{p} (3+p^2)}{C^p}$ for $C>2$, we obtain the  inequality \eqref{est-inf}.
\fin

\section{Reduced density matrices}
\label{sec:RDM}
In this section, we explain the notion of Wigner measures and its relationship with reduced density matrices. Most of the results we need are proved in \cite{AmNi1,AmNi3}, but for reader convenience we briefly recall them since they play an essential role in the proof of our main result.  The main observation is that reduced density matrices of a given sequence of  normal states have limit points with respect to the weak-$*$ topology  when  $n\to\infty$ and these limit points have a very particular structure. Actually, a non-commutative de Finetti theorem \cite{Stor} due to St{\o}rmer (motivated by classification of $C^*$-algebras and type factors) provides in some sense the structure of these limiting states. This is more apparent in the work of Hudson and Moody \cite{Hud,HuMo} where the authors focus on normal states which are also used in our setting. Actually, it turns out that with Wigner measures we can characterize the structure of the limit points more easily, without appealing to $C^*$-algebras formalism, and using probability measures in  more natural sets. Moreover, some compactness defect phenomena can be easily understood with the latter tool (see \cite{AmNi1,AmNi3}). More recently,  the authors Lewin, Nam and Rougerie   in \cite{LNR1} gave an alternative proof of the non-commutative de Finetti theorem (see also \cite{CHPS,LNR2} for application of this type of result).

\subsection{Wigner measures}
\label{sub.wigner}
In finite dimension, Wigner (or semi-classical) measures are well-known tools in the analysis of PDEs with particular scaling (see for instance \cite{PGe,GMMP,LiPa,Mar,Rob,Tar}). This idea was extended to the infinite dimensional case in \cite{AmNi1} and adapted to the framework of the mean field problem. Actually, the Borel probability measures $\mu_0$ appearing in  Proposition \ref{definetti} is what we call  Wigner measures of the sequence of density matrices $(\varrho_n)_{n\in\nz^*}$. This  concept is more general and one can deal with  arbitrary families  of normal states (or even trace class operators) on the Fock space. The main advantage is that we can identify these  measures $\mu_0$ by means of simpler quantities involving the Weyl operators (see Theorem \ref{th.wig-measure}) according to the formula:
\begin{equation*}
\lim_{n\to\infty} \Tr[\varrho_{n} W(\xi)]= \int_{\Z}
e^{i\sqrt{2} {\rm Re}\langle\xi, z\rangle} \, d\mu_0(z)\,, \quad\forall\xi\in\Z\,,
\end{equation*}
where $W(\xi)$ refers to the Weyl operator on the Fock space $\mathscr{H}$ with $\varepsilon=\frac{1}{n}$.
Therefore the mean field problem becomes a propagation problem of Wigner measures along the nonlinear flow of the (Hartree) equation \eqref{hartree-abs}. To enlighten the discussion let us consider a concret example. Let $\Psi_n=\varphi^{\otimes n}$ with $\varphi\in\Z$ and $ ||\varphi||_{\Z}=1$. It is easy to see that the $p$-reduced density matrices of $\varrho_n=|\Psi_n\ra\la\Psi_n|$ are $\varrho_n^{(p)}=|\varphi^{\otimes p}\ra\la
\varphi^{\otimes p}|$ and one can compute explicitly the Wigner measure of the sequence $(\varrho_n)_{n\in\nz}$  according to
Proposition \ref{definetti}:
$$
\lim_{n\to\infty}\Tr[\varrho_n^{(p)} B]=\la \varphi^{\otimes p},B\varphi^{\otimes p}\ra= \int_{\Z} \la z^{\otimes p} , B z^{\otimes p}\ra
 \; d\mu_0(z), \quad \mbox{ with }  \quad
\mu_0=\frac{1}{2\pi} \int_0^{2\pi} \delta_{e^{i\theta}\varphi} \, d\theta\,,
$$
where  $\delta_{e^{i\theta}\varphi}$  denotes the Dirac measure on  $\Z$  at the point $ e^{i\theta}\varphi$. So Theorem \ref{main-th} gives in particular the
convergence of the evolved reduced density matrices and in our example it yields
$$
\lim_{n\to\infty}\Tr[\varrho_n^{(p)}(t) B]= \int_{\Z}
\la z^{\otimes p} , B z^{\otimes p}\ra \; d\mu_t(z),\quad \mbox{ with }
\quad \mu_t=(\Phi_t)_\sharp\mu_0=\frac{1}{2\pi} \int_0^{2\pi} \delta_{e^{i\theta}\varphi_t} \, d\theta\,,
$$
where $\varphi_t$ is the solution of the nonlinear field (Hartree) equation  \eqref{hartree-abs} with initial condition $\varphi$. So, working with Wigner measures
allows to understand the superposition of states that may interact in the
mean field limit (see \cite{AmNi1}); and hence it provides a general and flexible point of view. We recall below the result that gives the construction of Wigner measures. It is a slight adaptation of \cite[Theorem 6.2]{AmNi1} including \cite[Lemma 2.14]{AmNi3}.

\begin{thm}
\label{th.wig-measure}
Let $(\varrho_n)_{n\in\nz^*}$ be a sequence of density matrices such that $\varrho_n\in \L^1(\vee^n\Z)$ for each $n\in\nz^*$. Then there exists a subsequence $({n_k})_{k\in\nz^*}$ and a Borel probability measure $\mu$ on $\Z$, called a Wigner measure,  such that for any $\xi\in\Z$,
\begin{equation}
\label{quan1}
\lim_{k\to\infty} \Tr[\varrho_{n_k} W(\xi)]= \int_{\Z}
e^{i\sqrt{2} {\rm Re}\langle\xi, z\rangle} \, d\mu(z)\,,
\end{equation}
with $W(\xi)$ referring  to the Weyl operator on the Fock space
$\mathscr{H}$ with the scaling  $\varepsilon=\frac{1}{n}\,$.
Moreover, the probability measure $\mu$ is $U(1)$ invariant and it is concentrated on the unit ball $B(0,1)$ of  the Hilbert space $\Z$ (i.e.: $\mu_0(B_{}(0,1))=1)$.
\end{thm}
The $U(1)$-invariance of the measure $\mu$ is a straightforward consequence of the fact that $\varrho_n\in\L^1(\vee^n\Z)$ for each $n\in\nz^*$. So, the above theorem says that the set of Wigner measures of a sequence $(\varrho_n)_{n\in\nz^*}$  is never empty and  we denote it by
$$
\M(\varrho_n, n\in\nz^*)\,.
$$
In practice and without loss of generality, one can assume in the analysis of the mean field problem that the set $\M(\varrho_n, n\in\nz^*)$ only contains a single measure.

\subsection{De Finetti Theorem}
 In this subsection we give the proof of Proposition \ref{definetti} which can be considered as a non-commutative de Finetti theorem.
 Moreover, the convergence \eqref{quan1} extends to Wick quantized symbols with compact kernels belonging to $\P^{\infty}$ and hence this proves the weak-* convergence of reduced density matrices. This result is proved in  \cite[Corollary 6.14]{AmNi1} and a slight adaptation  of it is recalled below.
\begin{prop}
 \label{pr.polycomp}
Let $(\varrho_{n})_{n\in\nz^*}$ be a sequence of density matrices such that $\varrho_n\in \L^1(\vee^n\Z)$ for each $n\in\nz^*$ and assume that $\M(\varrho_{n},
n\in\nz^*)=\left\{\mu_0\right\}$. Then the convergence
\begin{equation}
  \label{eq.convpolWW}
  \lim_{n\to\infty} \Tr[\varrho^{(p)}_{n} A] =\Tr[\varrho_{\infty}^{(p)} A]\,, \quad \mbox{ with }  \quad
\varrho_{\infty}^{(p)}=\int_{\Z} |z^{\otimes p}\rangle\langle z^{\otimes p}| \, d\mu_0(z)\,,
\end{equation}
holds for any $p\in\nz^*$ and any $A\in\L^{\infty}(\vee^p\Z)$.
\end{prop}

\bigskip
\noindent
{\bf Proof of Proposition \ref{definetti}}:\\
Suppose that for each  $p\in\nz^*$ and each compact
 operator $A\in\L^{\infty}(\vee^p\Z)$  the sequence $(\Tr[\varrho^{(p)}_{n} A])_{n\in\nz^*}$ converges. Then  there exist  trace-class operators $0\leq \varrho_\infty^{(p)}\leq 1$, $p\in\nz^*$, such that
 $$
 \lim_{n\to\infty} \Tr[\varrho^{(p)}_{n} A] =\Tr[\varrho_{\infty}^{(p)} A]\,,\quad \forall
 A\in\L^{\infty}(\Z)\,.
 $$
Let $\mu$ be any Wigner measure  in $\M(\varrho_{n},
n\in\nz^*)\neq \varnothing$. Then by Proposition \ref{pr.polycomp},  up to extraction of  subsequences, we see that
$$
\varrho_{\infty}^{(p)}=\int_{\Z} |z^{\otimes p}\rangle\langle z^{\otimes p}| \, d\mu(z)\,.
$$
So, this provides the existence of a Borel probability measure $\mu$ on $\Z$ with the appropriate properties. The uniqueness follows by \cite[Proposition 6.15]{AmNi1}.
\fin

\subsection{Defect of compactness}
\label{se.relywick}
The convergence in Proposition \ref{pr.polycomp} is with respect to the weak-* topology on
$\L^1(\vee^p \Z)$ which is the topological dual of $\L^\infty(\vee^p\Z))$ and the statement (\ref{eq.convpolWW}) does not hold in general for all  $A\in \L(\vee^p\Z)$, $p\in\nz^*$. Counterexamples exhibiting this  phenomenon of dimensional defect of compactness are given in \cite{AmNi1} (we call it in this way because of the similarity with finite dimension, although the source of defect here is the fact the phase-space is of infinite dimension and so bounded sets are not relatively compact in the norm topology). Actually,  the extension of  (\ref{eq.convpolWW}) to all bounded operators $A\in \L(\vee^p\Z)$ and  $p\in\nz^*$ depends on the sequence $(\varrho_{n})_{n\in\nz^*}$ and it turns out to be an important point in the mean field problem: we need this information when we take the limit $n\to\infty$ in the mean field expansion. Let $(\varrho_{n})_{n\in\nz^*}$ be a sequence of density matrices such that $\varrho_n\in \L^1(\vee^n\Z)$ for each $n\in\nz^*$. The  reduced density matrices  $(\varrho_{n}^{(p)})_{n\in\nz^*}$  weakly converges to   $\varrho_\infty^{(p)}\in\L^1(\vee^p\Z)$ if
\begin{eqnarray}
\label{weak-conv}
\lim_{n\to \infty} \Tr[\varrho_{n}^{(p)} \,A]=\Tr[\varrho_{\infty}^{(p)} \,A]\,, \quad
\;\forall A\in \L(\vee^p \Z)\,.
\end{eqnarray}
The following proposition provides a strong relationship between the
Wigner measures of a sequence of density matrices and the convergence of their reduced density matrices in the $\L^1$-norm topology.
\begin{prop}
\label{pr.wignergp0}
Let  $(\varrho_{n})_{n\in\nz^*}$  be a sequence of density matrices with $\varrho_n\in\L^1(\vee^n\Z)$ for each $n\in\nz^*$. Suppose that the reduced
density matrices $(\varrho^{(p)}_{n})_{n\in\nz^*}$ weakly converge to   $\varrho_\infty^{(p)}\in\L^1(\vee^p\Z)$ for each $p\in\nz^*$ according to \eqref{weak-conv}. Then there exists a unique $U(1)$-invariant Borel probability measure $\mu_0$  on $\Z$ such that for any $p\in\nz^*$,
\bean
\lim_{n\to\infty} ||\varrho^{(p)}_{n}-\varrho_{\infty}^{(p)}||_1=0\,, \quad \mbox{ with }  \quad
\varrho_{\infty}^{(p)}=\int_{\Z} |z^{\otimes p}\rangle\langle z^{\otimes p}| \, d\mu_0(z)\,.
\eean
Moreover, $\mu_0$ is the unique Wigner measure of $(\varrho_n)_{n\in\nz^*}$ and it is concentrated on the unit sphere $S_{}(0,1)$ of $\Z$ centred at the origin and of radius one (i.e.: $\mu_0(S_{}(0,1))=1)$.
\end{prop}
\proof
 The assumption on $(\varrho_n)_{n\in\nz^*}$  imply  that $\varrho^{(p)}_\infty$ are non-negative trace class operators with $\Tr[\varrho^{(p)}_\infty]=1$ and $\varrho^{(p)}_{n}$ converges  to $\varrho_{\infty}^{(p)}$ with respect to the weak topology in $\mathscr{L}^{1}(\vee^{p}\Z)$.
But since $\varrho^{(p)}_{n}$ and $\varrho^{(p)}_{\infty}$ are non negative trace-class operators with $\Tr[\varrho_n^{(p)}]=1=\Tr\left[\varrho^{(p)}_{\infty}\right]$, the $\L^1$-norm convergence follows according to \cite{Ak,DA,Sim}. In a more general framework, it is  said that $\L^{1}(\vee^{p}\Z)$ has the Kadec-Klee property (KK*) in the weak-* topology (see \cite{Len} and references therein). The (KK*) property on a dual Banach space means that the weak-* and norm convergence coincide on the unit sphere. \\
Thanks to the proof of Proposition \ref{definetti}, we know that $\mu_0$ is the unique Wigner measure of the sequence $(\varrho_n)_{n\in\nz^*}$. So, the measure $\mu_0$ is $U(1)$-invariant and it is concentrated on the unit ball of $\Z$ according to Theorem \ref{th.wig-measure}. Now, using the fact that $\Tr[\varrho_\infty^{(p)}]=1$, we get
$$
\int_{\Z} ||z||^{2p} \, d\mu_0(z)=1,\quad \forall p\in\nz^*\,.
$$
This easily yields that the measure is actually concentrated on the unit sphere.

\fin

\begin{cor}
  \label{pr.eqPPI}
  Let  $(\varrho_{n})_{n\in\nz^*}$  be a sequence of density matrices with $\varrho_n\in\L^1(\vee^n\Z)$ for each $n\in\nz^*$ and such that  $\mathscr{M}(\varrho_{n},n\in\nz^*)=\left\{\mu_0\right\}$. The two following conditions are equivalent:
$$
\left(\mu_0(S(1,0))=1\right)
\Leftrightarrow
\left(\forall p\in\nz^*,\;
\L^1-\lim_{n\to\infty} \varrho^{(p)}_{n}=\int_{\Z} |z^{\otimes p}\rangle\langle z^{\otimes p}| \, d\mu_0(z)\right)
$$
\end{cor}
\proof
Suppose that the Wigner measure $\mu_0$ is concentrated on the unit sphere, then by Proposition \ref{pr.polycomp} we see that
$(\varrho^{(p)}_{n})_{n\in\nz^*}$ weak-* converges to $\varrho^{(p)}_{\infty}$ which is a
non-negative trace-class operator with $\Tr[\varrho^{(p)}_{\infty}]=1$. So, again by the
Kadec-Klee property (KK*) of  $\L^1(\vee^p\Z)$ we obtain the $\L^1$-norm convergence for each $p\in\nz^*$.
\fin

\section{Rate of convergence}
\label{sec.rate}
In this section we give the proof of our main result  (Theorem \ref{main-th}). We start by proving
 an elementary estimate in Subsection \ref{sub.est} and then prove the result in Subsection \ref{sub.proof}.
\subsection{Preliminary estimate}
\label{sub.est}
Instead of estimating the quantities $||\varrho^{(p)}_n-\varrho_\infty^{(p)}||_1$
in the trace norm, we will work   essentially with
$$
\left| \Tr[\varrho_n A^{Wick} ]-\int_\Z \langle z^{\otimes p}, A\, z^{\otimes p}\rangle \,d\mu_0\right|\,.
$$
In that way, we can use the mean field expansion based on Wick calculus. The two quantities are comparable and this is given by the lemma below.
\begin{lem}
\label{lem.1}
Let $(\varrho_n)_{n\in\nz^*}$ be a sequence of density matrices such that $\varrho_n\in \L^1(\vee^p\Z)$ for each $n\in\nz^*$ and satisfying the assumptions of Theorem \ref{main-th}.  Then for any $n,p\in\nz^*, n\geq p$:
\bean
\left| ||\varrho^{(p)}_n-\varrho_\infty^{(p)}||_1- \sup_{A\neq 0}
\frac{| \Tr[\varrho_n A^{Wick} ]-\int_\Z \langle z^{\otimes p}, A\, z^{\otimes p}\rangle \,d\mu_0|}{||A||}\right|\leq
\frac{(p-1)^2}{n}\,,
\eean
with $\mu_0$ being the Wigner measure of $(\varrho_n)_{n\in\nz^*}$ and
$$
\varrho_\infty^{(p)}=\int_{\Z} |z^{\otimes p}\rangle\langle z^{\otimes p}| \, d\mu_0(z)\,.
$$
\end{lem}
\proof
For any $A\in\L(\vee^p\Z)$ and $\hbarr=\frac{1}{n}$:
$$
\Tr[\varrho_n \,A^{Wick} ]=\frac{n \cdots (n-p+1)}{n^p}  \,\Tr[\varrho_n^{(p)}  A ]\quad \mbox{ and } \quad
\Tr[\varrho_\infty^{(p)}A]=\int_{\Z} \langle z^{\otimes p}, A z^{\otimes p}\rangle\, d\mu_0(z)\,.
$$
 Hence, we get
\bean
\Big| \Tr[\varrho_n A^{Wick} ]-\int_\Z \langle z^{\otimes p}, A\, z^{\otimes p}\rangle \,d\mu_0\Big|&\leq&
\Big|1-\frac{n \cdots (n-p+1)}{n^p}\Big| \;||A||+ |\Tr[\varrho_n^{(p)}A]-\Tr[\varrho_\infty^{(p)}A]|\\
&\leq& \left[\Big|1-\frac{n \cdots (n-p+1)}{n^p}\Big| + ||\varrho^{(p)}_n-\varrho_\infty^{(p)}||_1\right] ||A||,
\eean
and also
\bean
| \Tr[(\varrho_n^{(p)} - \varrho_\infty^{(p)}) A]|&\leq&
\Big|1-\frac{n \cdots (n-p+1)}{n^p}\Big| \;||A||+ \Big|\Tr[\varrho_n A^{Wick}]-\int_\Z \langle z^{\otimes p}, A\, z^{\otimes p}\rangle \,d\mu_0\Big|\,.
\eean
The inequality
\bean
%\label{npfactor}
1-\frac{n \cdots (n-p+1)}{n^p}=1-\ds\prod_{j=1}^{p-1} \big(1-\frac j n\big) \leq 1-\bigg(1-\frac{p-1}{n}\bigg)^{p-1}\leq \frac{(p-1)^2}{n}\,,
\eean
gives the sought estimate.\fin

\subsection{Proof of the main theorem}
\label{sub.proof}
Recall that $\mu_t=(\Phi_t)_\sharp \mu_0$ in Theorem \ref{main-th}  with $\mu_0$ is the unique Wigner measure
of the sequence $(\varrho_n)_{n\in\nz^*}$ provided by Proposition \ref{definetti}.
\begin{lem}
\label{tech.lem.2}
For any $t\in\mathbb{R}$ such that $|t|<\frac{1}{8||\tilde Q||}$:
\begin{eqnarray}
\label{poisson-brack-conv}
\mu_t\big(\langle z^{\otimes p}, A z^{\otimes p}\rangle\big)=
\sum_{k=0}^{\infty} i^k \;\int_{0}^{t}d{t_{1}}\cdots\int_{0}^{t_{k-1}}dt_{k}\;\;
\mu_0\left( C^{(k)}_{0}(t_{k},\ldots,t_{1},t)\right)\,.
\end{eqnarray}
\end{lem}
\proof
We know already that the measure $\mu_0$ is concentrated in the ball  of radius $1$  centred at the origin according to Proposition \ref{definetti}. Hence, we deduce the inequality
$$
\left|\mu_0\left( C^{(k)}_{0}(t_{k},\ldots,t_{1},t)\right)\right | \leq \left\|\widetilde{C^{(k)}_{0}}(t_{k},\ldots,t_{1},t)\right\|_{\L(\vee^{p+k}\Z)}\,.
$$
The right hand side of \eqref{poisson-brack-conv} is absolutely convergent whenever $|t|<\frac{1}{8|| \tilde Q||}$ thanks to the estimate (ii) of Lemma \ref{tech.lem.1}:
\bean
\sum_{k=0}^{\infty} \;\int_{0}^{t}d{t_{1}}\cdots\int_{0}^{t_{k-1}}dt_{k}\;\;
\left|\mu_0\left( C^{(k)}_{0}(t_{k},\ldots,t_{1},t)\right)\right|\leq  2^{p-1} ||A|| \;
\sum_{k=0}^{\infty} \left(8  \;|t|\; ||\tilde Q||\;\right)^k.
\eean
Recall that according to \eqref{class-integ-form} the classical solution $t\mapsto z_t$ verifies for any $b\in\P_{p,p}$,
\bean
b(z_t)=b_t(w_0)+i\int_{0}^{t}\left\{Q_{t_1},b_t\right\}(w_{t_1})~dt_{1}\,,\quad \mbox{ with } \quad w_t=e^{i t\tilde h_0} z_t\,.
\eean
Iterating this formula, with $b(z)=\langle z^{\otimes p}, A z^{\otimes p}\rangle$, and using the absolute convergence checked above gives for all $||z||\leq 1$,
$$
\langle z_t^{\otimes p}, A z_t^{\otimes p}\rangle=\sum_{k=0}^{\infty}i^{k}\int_{0}^{t}d_{t_{1}}\cdots\int_{0}^{t_{k-1}}dt_{k}\;\;
C^{(k)}_{0}(t_{k},\ldots,t_{1},t;z)\,.
$$
Integrating with respect to $\mu_0$ and using the fact that $\mu_t=(\Phi_{t})_\sharp\mu_0$ yields \eqref{poisson-brack-conv}.
\fin

\bigskip
\noindent
{\bf Proof of Theorem \ref{main-th}:}\\
For reader convenience, we recall the assumptions of Theorem \ref{main-th} .
Let $(\alpha(n))_{n\in\nz^*}$ be a sequence of positive numbers with $\lim_{n\to\infty}\alpha(n)=\infty$ and
such that  $(\frac{\alpha(n)}{n})_{n\in\nz^*}$ is bounded.
Consider  $(\varrho_{n})_{n\in\nz^*}$ and $(\varrho_{\infty}^{(p)})_{p\in\nz^*}$ to be two sequences of density matrices with $\varrho_n\in\L^1(L^2_s(\rz^{dn}))$ and $\varrho_\infty^{(p)}\in\L^1(L^2_s(\rz^{dp}))$  for each $n,p\in\nz^*$.  Assume that there exist $C_0>0$, $C>2$, $\gamma\geq 1$ such that for all
$n,p\in\nz^*$ with $n\geq \gamma p$,
\bea
\label{init-ineq-bis}
\left\|\varrho^{(p)}_{n}-\varrho_{\infty}^{(p)}\right\|_1\leq
 C_0 \frac{C^p}{\alpha(n)}\,.
\eea
We first prove the estimate for short times and than extend it to arbitrary times. So, suppose that $|t|< \frac{1}{8C||\tilde Q||}$ with $C>2$ the constant provided by the main assumption. Thanks to Lemma \ref{lem.1} it is enough to estimate the quantity
$| \Tr[\varrho_n(t) \, A^{Wick} ]-\mu_t\big(\langle z^{\otimes p}, A\, z^{\otimes p}\rangle\big)|$ for any bounded
operator  $A\in\L(\vee^p\Z)$. So, the estimate in  Proposition  \ref{prop-3} yields
\bea
\label{eq.fes}
\left|{\rm Tr}[\varrho_{{n}} (t) \; A^{Wick}] -\mu_t\big(\langle z^{\otimes p}, A z^{\otimes p}\rangle\big)
\right| &\leq& C_0 \frac{C^{p}}{n} ||A|| +\mathfrak{R}(t)\,,
\eea
for some $C_0>0$ and
\bean
\mathfrak{R}(t)&=&
\left|\sum_{k=0}^{\infty}  i^k
 \int_0^t dt_1\cdots\int_0^{t_{k-1}} dt_k \;{\rm Tr}\left[\varrho_{n}\,
C_0^{(k)}(t_k,\cdots,t_1,t)^{Wick}\right]-\mu_t\big(\langle z^{\otimes p}, A z^{\otimes p}\rangle\big)\right| \\
&=&\left|\sum_{k=0}^{\infty}  i^k
 \int_0^t dt_1\cdots\int_0^{t_{k-1}} dt_k \; {\rm Tr}\left[\varrho_{n}
\,C_0^{(k)}(t_k,\cdots,t_1,t)^{Wick}\right]-\mu_0\big(C_0^{(k)}(t_k,\cdots,t_1,t)\big)\right|\,.
\eean
The last equality is a consequence of Lemma \ref{tech.lem.2}.
Using now the main assumption and the fact that $C_0^{(k)}(t_k,\cdots,t_1,t)$ are polynomials  in
$\P_{p+k,p+k}$, we get the inequality (we can assume that $t>0$ without loss of generality)
\bea
\label{t1}
\mathfrak{R}(t)&\leq& \sum_{k=0}^{\lfloor\frac{n}{\gamma}\rfloor-p}
 \int_0^t dt_1\cdots\int_0^{t_{k-1}} dt_k \;\left|{\rm Tr}\left[\varrho_{n}\,
C_0^{(k)}(t_k,\cdots,t_1,t)^{Wick}\right]-\mu_0\big(C_0^{(k)}(t_k,\cdots,t_1,t)\big)\right|\\ \label{t2}
&&+\sum^{n-p}_{k=\lfloor\frac{n}{\gamma}\rfloor-p+1}
 \int_0^t dt_1\cdots\int_0^{t_{k-1}} dt_k \;\left|{\rm Tr}\left[\varrho_{n}\,
C_0^{(k)}(t_k,\cdots,t_1,t)^{Wick}\right]-\mu_0\big(C_0^{(k)}(t_k,\cdots,t_1,t)\big)\right|\\  \label{t3}
&&+ \sum_{k=n-p+1}^{\infty}
 \int_0^t dt_1\cdots\int_0^{t_{k-1}} dt_k \;\left|\mu_0\big(C_0^{(k)}(t_k,\cdots,t_1,t)\big)\right|\,.
\eea
Thanks to  the estimate (ii) of Lemma \ref{tech.lem.1}, the right hand side of \eqref{t1}-\eqref{t3} is bounded by
\bean
\eqref{t1} &\leq& \frac{C^p}{\alpha(n)} \sum_{k=0}^{\lfloor\frac{n}{\gamma}\rfloor-p}   (8t C ||\tilde Q||)^k \;||A||\leq\frac{C^p}{\alpha(n)}  \frac{1}{1-8t C ||\tilde Q||}\;||A||\\
\eqref{t2} &\leq& 2\sum^{n-p}_{k=\lfloor\frac{n}{\gamma}\rfloor-p+1}  (8t ||\tilde Q||)^k \;||A||\leq 2\frac{(8t ||\tilde Q||)^{\lfloor\frac{n}{\gamma}\rfloor-p+1}}{1-8t ||\tilde Q||}\;||A||\\
\eqref{t3}& \leq& \sum_{k=n-p+1}^\infty (8t ||\tilde Q||)^k \;||A||=\frac{(8t ||\tilde Q||)^{n-p+1}}{1-8t ||\tilde Q||}\;||A||\,.
\eean
Since $|t|< \frac{1}{8C||\tilde Q||}$, we easily get  the bounds
\bean
\eqref{t2} \leq \frac{C^{p}}{C^{\lfloor\frac{n}{\gamma}\rfloor}}\;\frac{2||A||}{1-8t ||\tilde Q||}\leq \lambda \frac{C^{p}}{\alpha(n)}\;\frac{2||A||}{1-8t ||\tilde Q||} \quad \mbox{ and } \quad\eqref{t3} \leq \frac{C^{p}}{C^n}\;\frac{||A||}{1-8t ||\tilde Q||}\leq
\lambda\frac{C^{p}}{\alpha(n)}\;\frac{||A||}{1-8t ||\tilde Q||}\,,
\eean
 with $\lambda=\sup_{n\in\nz^*} \frac{\alpha(n)}{C^{\lfloor\frac{n}{\gamma}\rfloor}}$ which depends only on $C$ and the sequence
 $(\alpha(n))_{n\in\nz^*}$. Collecting theses estimates, we conclude that
 \bean
 \mathfrak{R}(t)&\leq& \frac{(1+3\lambda)}{1-8tC||\tilde Q||}\,
 \frac{C^p}{\alpha(n)}\, \,||A||\,.
 \eean
So, by Lemma \ref{lem.1} there exists $C_1>0$ such that
\bean
\left\|\varrho^{(p)}_{n}(t)-\varrho_{\infty}^{(p)}(t)\right\|_1\leq
 C_1 \frac{C^p}{\alpha(n)}\,,
\eean
uniformly in  time whenever $|t|\in [0,\frac{1}{16 C ||\tilde Q||}]$. Now, iterate the same reasoning
as much as needed to cover a time interval $[-T,T]$ with $T>0$ arbitrary.  Then one gets the existence of $C_T>0$ such that for all $t\in[-T,T]$,
\bean
\left\|\varrho^{(p)}_{n}(t)-\varrho_{\infty}^{(p)}(t)\right\|_1\leq
 C_T \frac{C^p}{\alpha(n)}\,.
\eean
\fin

\section{Examples and numerical simulations}
\label{sec.ENS}
In order to illustrate the main result of this article, it is useful to consider some examples and numerical simulations of states with an increasing degree of  correlation.   The notion of correlation is quite important in quantum information theory and it is related to the so-called quantum entanglement. So, there are several interesting examples of states in the latter field which are also useful for our purpose (Bell state, cat state, W state, GHZ state,...).

\subsection{Product states}
\begin{figure}[ht!]
  \centering
   \includegraphics[width=0.7\textwidth]{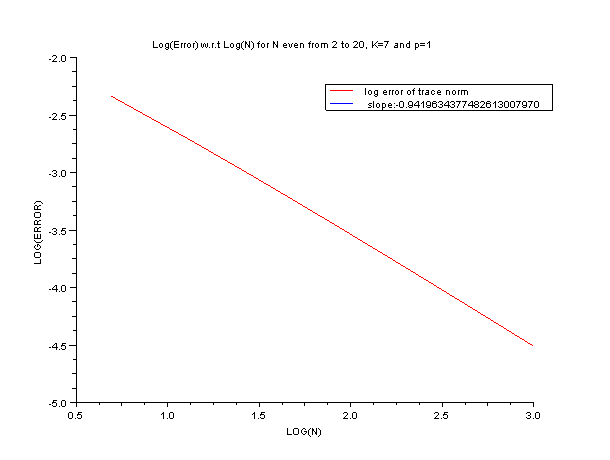}
  \caption{Log-log plot for factorized states}
  \label{fig:plot1}
\end{figure}

This is the most known example in mean field theory. It appears in the literature under the name of chaos, factorized, product or also Hermite states. It emphasizes the fact that, in the mean field limit, states  $\varphi^{\otimes n}$ that are prepared uncorrelated   will evolve into states which are close to be uncorrelated, namely $\varphi_t^{\otimes n}$ where $\varphi_t$ is a solution of the Hartree equation \eqref{hartree} with initial condition $\varphi$. It is easy to see that the factorized states
$$
\varrho_n=|\varphi^{\otimes n}\ra\la
\varphi^{\otimes n}|\,, \quad \mbox{ with } ||\varphi||=1\,,
 $$
 satisfy the assumption of Theorem \ref{main-th}. In fact, the $p$-reduced density matrices of $\varrho_n$ coincide with the limit
\bea
\label{prod}
 \varrho_n^{(p)}=|\varphi^{\otimes p}\ra\la
\varphi^{\otimes p}|=\varrho_\infty^{(p)}.
 \eea
 This means that in this example the rate of convergence at initial time $t=0$ is arbitrary fast. Remember that according to Theorem \ref{main-th} the $p$-particle reduced density matrix  $\varrho_{\infty}^{(p)}(t)$ is
\bean
\varrho_{\infty}^{(p)}(t)=\int_{\Z} |z^{\otimes p}\rangle\langle z^{\otimes p}| \, d\mu_t(z)=|\varphi_t^{\otimes p}\rangle\langle \varphi_t^{\otimes p}|\,\quad
\mbox{ with } \quad
\mu_t=\frac{1}{2\pi} \int_0^{2\pi} \delta_{e^{i\theta}\varphi_t} \, d\theta\,,
\eean
where $\varphi_t$ is a solution of the nonlinear field equation \eqref{hartree-abs}.
A numerical simulation performed on a discrete model (Figure \ref{fig:plot1}), shows that for the first marginal
\bean
\sup_{t\in \{t_1,\cdots,t_m\}\subset [0,1]}\mathrm{Log} \left\|\varrho^{(1)}_{n}(t)-\varrho_{\infty}^{(1)}(t)\right\|_1=-(1+\varepsilon )\mathrm{Log}(n) +  O(1)\,;
\eean
with a deviation $\varepsilon \simeq -0.06$ well within the expected computational inaccuracy. This is in very good agreement with our mathematical prevision, and indicates that the estimate in Theorem \ref{main-th} is not far from being optimal, in some sense.

\subsection{W states}
The W state is a multi-partite $n$-qubit entangled quantum state
$$
|W\rangle = \frac{1}{\sqrt{n}}(|100...0\rangle + |010...0\rangle + ... + |00...01\rangle)\,,
$$
where  $|1\rangle$ denotes a one particle excited state and $|0\rangle$  denotes the one particle ground state of two mode system. More generally, if $\Z$ is a  Hilbert space and   $\varphi,\psi$ are two normalized  orthogonal vectors in $\Z$ then
\bea
\label{def.Wst}
|W\rangle=\frac{1}{\sqrt{n}}(|\psi\otimes\varphi^{\otimes n-1}\rangle + |\varphi\otimes\psi\otimes\varphi^{\otimes n-2}\rangle + ... + |\varphi^{\otimes n-1}\otimes\psi\rangle)\,.
\eea
\begin{lem}
Let $(\varrho_n)_{n\in\nz^*}$ denotes a sequence of W states as in \eqref{def.Wst}. Then
for all $n,p\in\nz^*$ such that $n\geq p$:
$$
||\varrho_n^{(p)}-\varrho_\infty^{(p)}||_1\leq 2\,\frac{p}{n}\, \quad \mbox{ with }
\quad \varrho_\infty^{(p)}=|\varphi^{\otimes p}\ra\la
\varphi^{\otimes p}|\,.
$$
\end{lem}
\proof
A simple computation yields for any $A\in\L(\vee^p\Z)$:
\bean
\la W, A\otimes 1^{\otimes(n-p)} W\ra=\frac{n-p}{n} \la\varphi^{\otimes p} , A \varphi^{\otimes p}\ra+\frac{p}{n} \la W_p, A W_p\ra\,,
\eean
where $|W_p\ra=\frac{1}{\sqrt{p}}\big(\psi\otimes\varphi^{\otimes (p-1)}+\cdots+
\varphi^{\otimes (p-1)}\otimes \psi\big)$.
So that the $p$-reduced density matrices of $\varrho_n$ is
\bean
\varrho_n^{(p)}&=&
\frac{n-p}{n} \;|\varphi^{\otimes p}\ra\la\varphi^{\otimes p}|+\frac{p}{n}\; |W_p\ra\la W_p|\\
&=&\frac{n-p}{n}\;\varrho_\infty^{(p)}+\frac{p}{n} \;|W_p\ra\la W_p|\,,
\eean
Hence the estimate follows since $W_p$ is a normalized vector.
\fin

\subsection{GHZ states}
 The GHZ (Greenberger-Horne-Zeilinger) state is a multipartite  entangled quantum state. In a two-mode system it is given by the formula
 $$
|\mathrm{GHZ}\rangle = \frac{|0\rangle^{\otimes n} + |1\rangle^{\otimes n}}{\sqrt{2}}\,,
$$
So, it can be generalized as follows
\bea
\label{def.GHZ}
|\mathrm{GHZ}\rangle = \frac{|\varphi\rangle^{\otimes n} + |\psi\rangle^{\otimes n}}{\sqrt{2}}\,,
\eea
 where $\varphi$ and $\psi$ are two normalized (orthogonal) vectors in a given Hilbert space $\Z$. So, the $n$-partite GHZ states are superposition of uncorrelated states and it is again easy to  check that their $p$-reduced density matrices coincide with their limit as in \eqref{prod}. Hence, Theorem \ref{main-th} provides a rate of convergence for this example too with $\alpha(n)=n$ rate.

\subsection{Twin states}
Let $\varphi_1,\varphi_2\in\Z$ be two normalized orthogonal vectors. The  \emph{twin states} are rank one projectors $\varrho_n=|\Psi_n\ra\la\Psi_n|$  given by
\bea
\label{twin-vec}
\Psi_n=\sqrt{\frac{n!}{n_1!n_2!}} \,\S_n \varphi_1^{\otimes n_1}\otimes \varphi_2^{\otimes n_2}\,,
\eea
 with $n=n_1+n_2$ and $n_1=n_2\in\nz^*$. This sequence of states have a unique  Wigner measure $\mu$ computed in \cite{AmNi3}. So, after identification of the Hilbert space $\Z$ as $\mathbb{C}\varphi_1\times\mathbb{C}\varphi_2\times \Z_1^\perp$, with $\Z_1^\perp$ the orthogonal subspace to $\mathbb{C}\varphi_1\oplus\mathbb{C}\varphi_2$, the measure $\mu$ reads
\bea
\label{twin-mes}
\mu=\delta^{S^1}_{\frac{\varphi_1}{\sqrt{2}}}\otimes\delta^{S^1}_{\frac{\varphi_2}{\sqrt{2}}}\otimes
\delta_0^{\perp}\, \quad \mbox{ with } \quad \delta^{S^1}_{\frac{\varphi_j}{\sqrt{2}}}=\frac{1}{2\pi} \int_0^{2\pi} \delta_{e^{i\theta}\frac{\varphi_j}{\sqrt{2}}} \, d\theta\,, \; j=1,2.
\eea
Remark that in this example the measure $\mu_t=(\Phi_t)_\sharp\mu$ is  quite correlated because of the nonlinear  effect of the flow and  the situation differs significantly from the simple picture of uncorrelated states ( here $\Phi_t$ is the flow of the nonlinear field equation \eqref{hartree-abs} ).

\begin{lem}
\label{twin}
Let $(\varrho_n)_{n\in\nz^*}$ be a sequence of twin states with $\mu$ its Wigner measure given in \eqref{twin-mes}. Then for any $n,p\in\nz^*$ such that $n\geq 2 p$:
\bean
\left\|\varrho_n^{(p)}-\varrho_\infty^{(p)}\right\|_1 \leq  2^p \frac{p^2}{n-p} \quad \mbox{ with } \quad
\varrho_\infty^{(p)}=\int_\Z |z^{\otimes p}\ra\la z^{\otimes p}|  \, d\mu \,.
\eean
\end{lem}
\proof
Let $\Psi_n$ be the vector  given by \eqref{twin-vec}.   A simple computation yields
\bean
\la \Psi_n, A\otimes 1^{\otimes (n-p)} \,\Psi_n\ra&=&\frac{n!}{n_1!n_2!} \frac{1}{(n!)^2} \sum_{\sigma,\pi\in\mathfrak{S}(n)}
\la T_\sigma \varphi_1^{\otimes n_1}\otimes \varphi_2^{\otimes n_2}, A\otimes 1^{\otimes (n-p)}  \,T_\pi \varphi_1^{\otimes n_1}\otimes \varphi_2^{\otimes n_2}\ra,
\eean
where $T_\sigma$ denotes the operator on $\otimes^n\Z$ defined for any $\sigma\in\mathfrak{S}(n)$ by
$$
T_\sigma f_1\otimes \cdots\otimes f_n=f_{\sigma_1}\otimes \cdots\otimes f_{\sigma_n}\,.
$$
For $m,n\in\nz$, $k\leq m$, we denote
$$
\mathcal{I}_m^{(k)}=\left\{ i:\{1,\cdots,m\}\to\{1,2\},   \sharp i^{-1}(\{1\})=k\right\}\,.
$$
So, there is a correspondence between permutations $\sigma\in\Sigma_n$ and maps $i\in\mathcal{I}_n^{(n_1)}$ according to
\bea
\label{sigi}
T_\sigma  \varphi_1^{\otimes n_1}\otimes \varphi_2^{\otimes n_2}= \varphi_{i(1)}\otimes\cdots \otimes
\varphi_{i(n)}=:\varphi(i)\,.
\eea
Since the cardinal of the set of $\sigma\in\Sigma(n)$ such that \eqref{sigi} holds for the same $i\in \mathcal{I}_n^{(n_1)}$  is
equal to $n_1!n_2!\,$, we see that
\bean
\la \Psi_n, A\otimes 1^{\otimes (n-p)} \,\Psi_n\ra&=&\frac{n!}{n_1!n_2!} \bigg(\frac{n_1! n_2!}{n!}\bigg)^2
\; \sum_{i,j\in \mathcal{I}_n^{(n_1)}}
\la \varphi(i), A\otimes 1^{\otimes (n-p)}  \,\varphi(j)\ra\,.
\eean
In the above sum if $i\neq j$ on the set $\{p+1,\cdots,n\}$ then the scalar product is null because of the orthogonality condition on
the vectors $\varphi_1,\varphi_2$.  So this simplifies the sum and actually we can decompose it according to the number of occurrence of $\varphi_1$ in the first $p$ vectors constituting  $\varphi(i)$, i.e.:
\bean
\mathcal{I}_n^{(n_1)}=\ds\cup_{k=0}^{p} \left\{ i\in  \mathcal{I}_n^{(n_1)}, \sharp i^{-1}(\{1\}) \cap
\{1,\cdots,p\}=k\right\}\,=:\cup_{k=0}^{p} \mathcal{I}_{n,k}^{(n_1)} .
\eean
Hence,
\bean
\la \Psi_n, A\otimes 1^{\otimes (n-p)} \,\Psi_n\ra=\frac{n_1! n_2!}{n!} \sum_{k=0}^{p}
\; \sum_{i\in \mathcal{I}_{n,k}^{(n_1)}}  \; \sum_{j\in \mathcal{I}_{n}^{(n_1)}, j=i_{| p+1,\cdots,n}}
\la \varphi(i), A\otimes 1^{\otimes (n-p)}  \,\varphi(j)\ra\,.
\eean
If we fix the first $p$ values of $i$ and $j$ and variate the $(n-p)$ others then the scalar product $\la \varphi(i), A\otimes 1^{\otimes (n-p)}  \,\varphi(j)\ra$ will not change as long as $j=i_{| \{p+1,\cdots,n\}}$. Actually, there is $C_{n-p}^{n_1-k}$
configurations for each choice of $i(1),\cdots,i(p),j(1),\cdots,j(p)$ such that $\sharp i^{-1}(\{1\}) \cap
\{1,\cdots,p\}=\sharp j^{-1}(\{1\}) \cap
\{1,\cdots,p\}=k$ . Hence, we get
\bean
\la \Psi_n, A\otimes 1^{\otimes (n-p)} \,\Psi_n\ra=\frac{n_1! n_2!}{n!} \sum_{k=0}^{p}  \; C_{n-p}^{n_1-k} \;
\; \sum_{i\in \mathcal{I}_{p}^{(k)}}  \;\sum_{j\in \mathcal{I}_{p}^{(k)}}
\la \varphi(i) , A \,\varphi(j)\ra\,.
\eean
Observe that for all $0\leq k\leq p$ and $2p\leq n$:
$$
\lim_{n\to \infty} \frac{C_{n-p}^{n_1-k}}{C_n^{n_1}}=\frac{1}{2^p}\,.
$$
So, we see that the limit of the $p$-reduced density matrices is
$$
\varrho_\infty^{(p)}=\frac{1}{2^{p}}\sum_{k=0}^{p}  \;
\; |\psi_k\left\rangle \;\right\langle\psi_k\big|=\int_\Z |z^{\otimes p}\ra\la z^{\otimes p}|  \, d\mu \,, \quad \mbox{ with } \quad \psi_k=\sum_{i\in \mathcal{I}_{p}^{(k)}} \varphi(i)\,,
$$
where $\mu$ is the Wigner measure of the sequence $(\varrho_n)_{n\in\nz^{*}}$ given in \eqref{twin-mes}. In particular, the orthogonality of the family $(\psi_k)_{1,\cdots,p}$ gives
\bean
1=||\varrho_\infty^{(p)}||_1=\frac{1}{2^{p}}\sum_{k=0}^{p}  \;
\; \big\|\psi_k\big\|^2
\eean
Therefore a simple estimate yields
\bean
\left\|\varrho_n^{(p)}-\varrho_\infty^{(p)}\right\|_1&\leq& \max_{k=1,\cdots,p} \left|
1-2^{p}\frac{C_{n-p}^{n_1-k}}{C_n^{n_1}}\right| \; \sum_{k=0}^p \frac{1}{2^{p}}
 ||\psi_k||^2 \\
&\leq&   \max_{k=1,\cdots,p} \left|
1-2^{p}\frac{C_{n-p}^{n_1-k}}{C_n^{n_1}}\right| \,.
\eean
So, the result follows once we  prove
\bean
\max_{k=1,\cdots,p} \left|
1-2^{p}\frac{C_{n-p}^{n_1-k}}{C_n^{n_1}}\right| \leq 2^p \frac{p^2}{n-p}\;.
\eean
In fact, for any $(a_i)_{1,\cdots,r}$ such that $0\leq a_i\leq 1$ the following simple estimates hold true
\bea
\label{est-el1}
0\leq &1-\prod_{i=1}^r (1-a_i)&\leq r \;\max_{1,\cdots,r} a_i\\ \label{est-el2}
 0\leq &\prod_{i=1}^r (1+a_i)-1&\leq 2^{r-1} r \;\max_{1,\cdots,r} a_i\,.
\eea
By writing
\bean
2^{p}\frac{C_{n-p}^{n_1-k}}{C_n^{n_1}}=\underbrace{\prod_{i=1}^{k-1} \big(1-\frac{i}{n-i}\big) \times
\prod_{j=k+1}^{p-k-1} \big(1-\frac{j-k}{n-k-j}\big)}_{T_1} \times \overbrace{\prod_{s=0}^{\min(p-k-1,k-1)} \big(1+\frac{k-s}{n-k-s}\big)}^{T_2}
\eean
we can see that the product $T_1=\prod_{i=1}^p (1-\beta_i)$ while the last one is $T_2=\prod_{j=1}^{p} (1+\gamma_j)$ with
 $0\leq\beta_i,\gamma_j\leq 1$ (some of the $\beta_i,\gamma_j$ are null). Hence applying \eqref{est-el1}-\eqref{est-el2}, we obtain
\begin{samepage}
 \bean
 |1-T_1 T_2|&\leq& |T_1| \,(T_2-1)+(1-T_1)\\
 &\leq& 2^{p-1} p\,\max_{1,\dots,p} \gamma_j + p \max_{1,\dots,p} \beta_i \\
 &\leq& 2^{p-1} p\,\frac{p}{n-p}+p \frac{p}{n-p}\,.
\eean
\fin
\end{samepage}

\begin{figure}[ht!]
  \centering
   \includegraphics[width=0.7\textwidth]{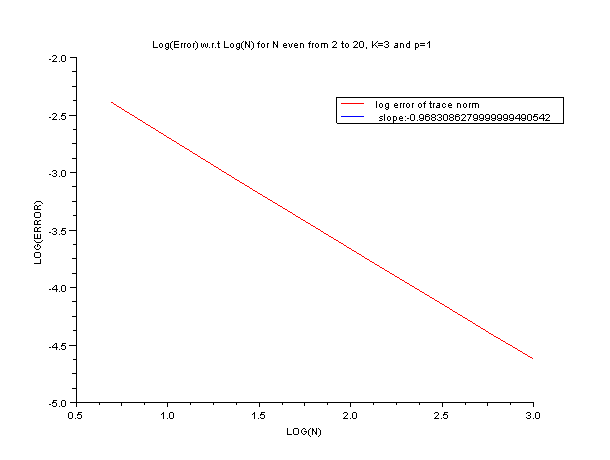}
  \caption{Log-log plot for twin states}
  \label{fig:plot2}
\end{figure}

Again in this example, a numerical simulation indicates a $1/n$ order of convergence for the first reduced density matrix (Figure \ref{fig:plot2}).

\medskip

Finally, we bring to reader's attention the fact that any rate of convergence is actually possible. In fact take the following example
$$
\varrho_n=\Big(1-\frac{1}{\alpha(n)} \Big)\,|e_1^{\otimes n}\ra\la e_1^{\otimes n}|+ \frac{1}{\alpha(n)}  \,|e_2^{\otimes n}\ra\la e_2^{\otimes n}|\,,
$$
with $(\alpha(n))_{n\in\nz^*}$ such that $\alpha(n)\geq 1$, $\alpha(n)\to \infty$ and $e_1,e_2$ are two normalized orthogonal vectors.   So, it is easy to see that
$$
\varrho_n^{(p)}=\Big(1-\frac{1}{\alpha(n)} \Big)\,|e_1^{\otimes p}\ra\la e_1^{\otimes p}|+ \frac{1}{\alpha(n)}  \,|e_2^{\otimes p}\ra\la e_2^{\otimes p}| \quad \mbox{ and } \quad \varrho_\infty^{(p)}=\,|e_1^{\otimes p}\ra\la e_1^{\otimes p}|\,.
$$
Therefore, for each $p\in\nz^*$, the following equality is satisfied:
$$
\left\|\varrho_n^{(p)}-\varrho_\infty^{(p)}\right\|_1 = \frac{2}{\alpha(n)}  \,.
$$
\medskip

\section*{Acknowledgement}
The research of the second and third  authors  has been supported respectively  by the Centre Henri Lebesgue ANR-11-LABX-0020-01 and ANR-11-IS01-0003 Lodiquas.

\bibliographystyle{empty}

\begin{thebibliography}{999}
\bibitem{Ak} A.~C.~Akemann.
\newblock The dual space of an operator algebra.
\newblock Trans. Amer. Math. Soc. 126, (1967) 286--302.

\bibitem{AmBr}
Z.~Ammari and S.~Breteaux.
\newblock Propagation of chaos for many-boson systems in one dimension with a
  point pair-interaction.
\newblock {\em Asymptot. Anal.}, 76(3-4):123--170, 2012.

\bibitem{AmNi1} Z.~Ammari, F.~Nier.
\newblock Mean field limit for bosons and infinite dimensional phase-space analysis.
\newblock Ann.~Henri Poincar{\'e} 9 (2008), 1503--1574.

\bibitem{AmNi2} Z.~Ammari, F.~Nier.
\newblock Mean field limit for bosons and propagation of Wigner measures.
\newblock J.~Math.~Phys. 50 (2009).

\bibitem{AmNi3} Z.~Ammari, F.~Nier.
\newblock Mean field propagation of Wigner measures and BBGKY hierarchies for general bosonic states.
\newblock J.~Math.~Pures Appl. 95 (2011), 585--626.

\bibitem{AmNi4} Z.~Ammari and F.~Nier.
\newblock {Mean field propagation of infinite dimensional Wigner measures with
  a singular two-body interaction potential}.
\newblock To appear in Ann. Sc. Norm. Super. Pisa Cl. Sci.

\bibitem{Ana} I.~Anapolitanos.
\newblock Rate of Convergence Towards the Hartree von Neumann Limit in the Mean-Field Regime,
\newblock Lett Math Phys  98 (2011), 1--31.

\bibitem{BGM} C.~Bardos, F.~Golse, N.~Mauser.
\newblock Weak coupling limit of the n-particle Schr{\"o}dinger equation.
\newblock Methods Appl. Anal. 7 (2000), 275--293.

\bibitem{BEGMY} C.~Bardos, L.~Erd{\"o}s, F.~Golse, N.~Mauser, H-T.~Yau.
\newblock Derivation of the Schr{\"o}dinger-Poisson equation from the quantum N-body problem.
\newblock C.R. Math. Acad. Sci. Paris 334 (2002), 515--520.

\bibitem{Ber} F.A.~Berezin.
\newblock \textit{The method of second quantization.}
\newblock Second edition. ``Nauka'', Moscow, (1986).

\bibitem{BoLe} J.M.~Bony, N.~Lerner.
\newblock Quantification asymptotique et microlocalisation d'ordre
  sup{{\'e}}rieur {I}.
\newblock Ann. Scient. Ec. Norm. Sup., $4^{e}$ s{{\'e}}rie 22 (1989), 377--433.

\bibitem{C} X.~Chen.
\newblock Second order corrections to mean field evolution for weakly
   interacting bosons in the case of three-body interactions.
\newblock Arch. Ration. Mech. Anal. 203, (2012), no 2, 455--497.

\bibitem{CLS} L.~Chen, J.O.~Lee, B.~Schlein.
\newblock Rate of convergence towards Hartree dynamics.
\newblock J. Stat. Phys. 144, (2011), No. 4, 872--903.

\bibitem{ChPa} T.~Chen, N.~Pavlovi\'c.
\newblock The quintic NLS as the mean field limit of a boson gas with three-body interactions.
\newblock J. Funct. Anal. 260 (2011), no. 4, 959--997.

\bibitem{CHPS}   T. Chen, C. Hainzl, N. Pavlovi{\'c}, R. Seiringer.
\newblock On the well-posedness and scattering for the Gross-Pitaevskii
   hierarchy via quantum de Finetti.
   \newblock Lett. Math. Phys. 104, (2014), 871--891.

\bibitem{DA} G.~F.~Dell'Antonio.
\newblock On the limits of sequences of normal states.
\newblock  Comm. Pure Appl. Math., 20 (1967),  413--429.

\bibitem{DG} J.~Derezi{\'n}ski, C.~G\'erard.
\newblock \textit{Mathematics of quantization and quantum fields}.
 \newblock Cambridge Monographs on Mathematical Physics,   Cambridge University Press,  (2013).


\bibitem{ElSc} A. Elgart, B. Schlein.
\newblock Mean field dynamics of boson stars
\newblock Comm. Pure and Appl. Math. Vol. 60, (2005) 500--545.

\bibitem{ErYa} L.~Erd{\"o}s, H.T.~Yau.
\newblock Derivation of the nonlinear Schr{\"o}dinger equation from a
many body Coulomb system.
\newblock Adv. Theor. Math. Phys. 5 (2001), 1169--2005.

\bibitem{ESY1} L.~Erd{\"o}s, B.~Schlein, H.T.~Yau.
\newblock Derivation of the cubic non-linear Schr{\"o}dinger equation from quantum dynamics of many-body systems.
\newblock Invent. Math. 167 no. 3  (2007), 515--614.

\bibitem{ESY2} L.~Erd{\"o}s, B.~Schlein, H.T.~Yau.
\newblock Derivation of the Gross-Pitaevskii equation for the dynamics of Bose-Einstein condensate.
\newblock Ann. of Math. (2) 172 (2010), no. 1, 291-370.

\bibitem{F} M.~Falconi.
\newblock Mean field limit of bosonic systems in partially factorized states and their linear combinations.
\newblock Arxiv http://fr.arxiv.org/abs/1305.5699.

\bibitem{FGS} J.~Fr{\"o}hlich, S.~Graffi, S.~Schwarz.
\newblock Mean-field- and classical limit of many-body
Schr{\"o}dinger dynamics for bosons.
\newblock Comm. Math. Phys.  271, No. 3 (2007), 681--697.

\bibitem{FKP} J.~Fr{\"o}hlich, A.~Knowles, A.~Pizzo.
\newblock Atomism and quantization.
\newblock  J. Phys. A 40, no. 12 (2007), 3033--3045.

\bibitem{FKS} J.~Fr{\"o}hlich, A.~Knowles, S.~Schwarz.
\newblock On the Mean-field limit of bosons with Coulomb two-body interaction
\newblock Comm. Math. Phys. 288, No. 3 (2009), 1023--1059.

\bibitem{PGe} P.~G{\'e}rard.
\newblock Microlocal defect measures.
\newblock Comm. Partial Differential Equations 16 (1991), no. 11, 1761-1794.

\bibitem{GMMP} P.~G{\'e}rard, P.A.~Markowich, N.J.~Mauser, F.~Poupaud.
\newblock Homogenization limits and Wigner transforms.
\newblock Comm. Pure Appl. Math. 50 no. 4 (1997), 323--379.

\bibitem{GiVe1} J.~Ginibre, G.~Velo.
\newblock The classical field limit of scattering theory for nonrelativistic
many-boson systems. I.
\newblock Comm. Math. Phys. 66 (1979), 37--76.

\bibitem{GiVe2} J.~Ginibre, G.~Velo.
\newblock The classical field limit of scattering theory for nonrelativistic
many-boson systems. II.
\newblock Comm. Math. Phys. 68, (1979), 45--68.

\bibitem{GMM} M.~Grillakis, M.~Machedon, D.~Margetis.
\newblock Second-order corrections to mean field evolution of weakly
   interacting bosons. I.
   \newblock Comm. Math. Phys. 294, (2010), no 1, 273--301.

\bibitem{GMP} S.~Graffi, A.~Martinez, M.~Pulvirenti.
\newblock Mean-field approximation of quantum systems and classical limit.
\newblock Math. Models Methods Appl. Sci. 13 No. 1 (2003), 59--73.

\bibitem{Hep} K.~Hepp.
\newblock The classical limit for quantum  mechanical correlation functions.
\newblock  Comm.~Math.~Phys. 35 (1974), 265--277.

\bibitem{Hud} R.~L. Hudson.
\newblock Analogs of de {F}inetti's theorem and interpretative problems of
  quantum mechanics.
\newblock {\em Found. Phys.}, 11(9-10):805--808, 1981.

\bibitem{HuMo} R.~L. Hudson and G.~R. Moody.
\newblock Locally normal symmetric states and an analogue of de {F}inetti's
  theorem.
\newblock {\em Z. Wahrscheinlichkeitstheorie und Verw. Gebiete},
  33(4):343--351, 1975/76.

\bibitem{KlMa} S.~Klainerman, M.~Machedon.
\newblock On the uniqueness of solutions to the Gross-Pitaevskii hierarchy.
\newblock Comm. Math. Phys. 279, (2008).

\bibitem{KP} A.~Knowles, P.~Pickl.
\newblock Mean-field dynamics: singular potentials and rate of convergence.
\newblock Comm. Math. Phys. 298 (2010), 101--138.

\bibitem{Len} C.~J.~Lennard.
\newblock $\mathcal{C}_1$ is uniformly Kadec-Klee.
\newblock Proc. Amer. Math. Soc. 109 (1990), 71--77.

\bibitem{LNR2} M.~Lewin, P.T.~Nam, N.~Rougerie.
\newblock Remarks on the quantum de Finetti theorem for bosonic systems.
\newblock Appl. Math. Res. Express (AMRX), in press, 2014.

\bibitem{LNR1} M.~Lewin, P.T.~Nam, N.~Rougerie.
\newblock Derivation of Hartree's theory for generic mean-field Bose gases.
\newblock Adv. Math., 254, (2014), 570--621.

\bibitem{QB} Q.~Liard, B.~Pawilowski.
\newblock Mean field limit for bosons with compact kernels interactions by Wigner measures transportation.
\newblock J. Math. Phys. 55, 092304 (2014).

\bibitem{LSSY}  E.H.~Lieb, R.~Seiringer,  J.P.~Solovej, J.~Yngvason.
\newblock \textit{The mathematics of the Bose gas and its condensation.}
\newblock Birkh{\"a}user (2005).

\bibitem{LiPa} P.L.~Lions, T.~Paul.
\newblock Sur les mesures de Wigner.
\newblock Rev. Mat. Iberoamericana 9 no. 3 (1993), 553--618.

\bibitem{Mar} A.~Martinez.
\newblock \textit{An Introduction to Semiclassical
  Analysis and Microlocal Analysis.}
\newblock Universitext, Springer-Verlag, (2002).

\bibitem{Pi} P.~Pickl.
\newblock A simple derivation of mean field limits for quantum systems.
\newblock Lett. Math. Phys. 97 (2011) 151--164.

\bibitem{Sim} B.~Simon.
\newblock Trace ideals and their applications.
\newblock Second edition. Mathematical Surveys and Monographs, 120. AMS, Providence, RI, 2005.

\bibitem{Rob} D.~Robert.
\newblock \textit{Autour de l'approximation semi-classique.}
\newblock Progress in Mathematics, 68.
 Birkh{\"a}user  (1987).

\bibitem{RoSc} I.~Rodnianski, B.~Schlein.
\newblock Quantum Fluctuations and Rate of Convergence towards Mean Field Dynamics.
\newblock Comm. Math. Phys. 291, No 1 (2009), 31--61.

\bibitem{Spo} H.~Spohn.
\newblock Kinetic equations from Hamiltonian dynamics.
\newblock Rev. Mod. Phys. 52, No. 3 (1980), 569--615.

\bibitem{Stor}
E.~St{\o}rmer.
\newblock Symmetric states of infinite tensor products of {$C^{\ast}
  $}-algebras.
\newblock {\em J. Functional Analysis}, 3:48--68, 1969.

\bibitem{Tar} L.~Tartar.
\newblock  H-Measures, a New Approach for Studying Homogenization.
\newblock  Oscillations and Concentration Effects in Partial Differential Equations, Proceedings of the Royal Society Edinburgh, 115-A (1990), 193--230.

\end{thebibliography}

\end{document}